\documentclass[]{spie}  
\usepackage[]{graphicx}
\usepackage{amsmath,amssymb}

\newcommand{\p}{\partial}
\title{Merging Nonlinear Optics and Negative-Index Metamaterials}

\author{Alexander K. Popov\supit{a} and Vladimir M. Shalaev\supit{b}
\skiplinehalf
\supit{a}Department of Physics and Astronomy, University of Wisconsin-Stevens Point,\\  Stevens Point, WI 54481, U. S. A. \\
\supit{b}Birck Nanotechnology Center and School of Electrical and Computer Engineering,\\ Purdue University, West Lafayette, IN 47907, U. S. A.
}

\authorinfo{Send correspondence to Alexander Popov: E-mail: apopov@uwsp.edu
}

\begin{document}
\maketitle

\begin{abstract}
The extraordinary properties of nonlinear optical propagation processes in double-domain positive/negative index metamaterials are reviewed. These processes include second harmonic generation, three- and four-wave frequency mixing, and optical parametric amplification. Striking contrasts with the properties of the counterparts in ordinary materials are shown. We also discuss the possibilities for compensating strong losses inherent to plasmonic metamaterials, which present a major obstacle in numerous exciting applications, and the possibilities for creation of unique ultracompact photonic devices such as data processing chips and nonlinear-optical sensors. Finally, we propose similar extraordinary three-wave mixing processes in crystals based on optical phonons with negative dispersion.
\end{abstract}



\section{INTRODUCTION}
\label{sec:intro}  

In the late 1960s, V. G. Veselago considered the propagation of electromagnetic waves in an fictitious, isotropic medium with simultaneously negative dielectric permittivity $\epsilon$ and magnetic permeability $\mu$ and showed that it would exhibit very unusual properties \cite{Vesel1,Vesel2}.
Specifically, the simultaneously negative dielectric permittivity,
$\epsilon<0$, and magnetic permeability, $\mu<0$, would lead to a negative
refraction index and to a left-handed triplet of the electric
field, magnetic field and the wavevector. The energy flow (Poynting
vector) in this case appears counter-directed with respect to the
wavevector. This is rather counter-intuitive and in a sharp contrast
with normal, positive-index materials (PIMs), also referred to as
right-handed materials (RHMs). Negative-index materials (NIMs), also referred to as left-handed materials (LHMs), do not exist naturally. It was also generally accepted in physical optics that the magnetization at optical frequencies is negligible and, hence, did not play any essential role \cite{Land}. In accordance with
this, the magnetic permeability $\mu$ was normally set to be equal to one in the basic Maxwell's equations describing linear and nonlinear optical processes.

Metamaterials, i.e., artificially designed and engineered materials, can have properties unattainable in nature, including a negative refractive index. As outlined above, the \emph{backwardness} of electromagnetic waves, i.e., the phenomenon of counter-directed energy flow and phase velocity for electromagnetic waves, does not exist in naturally occurring materials. Its realization has become feasible owing to the advent of nanotechnologies.  Significant progress in the design of bulk metamaterials has been achieved during the last few years (see, e.g., \cite{Jas,SM,Zh}). This dictates a revision of many concepts concerning electromagnetic propagation processes in man-made materials, which hold the promise of revolutionary breakthroughs in microwave and photonic device technologies \cite{Sh}.

Negative refraction can be implemented to develop a wide variety of devices with enhanced and uncommon functions, such as hyper-resolution lenses and optical cloaking devices. Nanostructured metamaterials are  expected to play a key role in the development of all-optical data processing chips. The opposite directions of the energy flow and wavevector (phase velocity) in NIMs determines unique linear and nonlinear optical processes. Initially, metamaterials with a magnetic response and a negative refractive index were fabricated for the microwave and terahertz frequency ranges \cite{Smith1,Smith11,Smith2}. The optical frequency range imposes increasing difficulties and challenges for metamaterials. A negative magnetic permeability in the optical range, which is a
precursor for a negative refraction, has been demonstrated in Refs. \cite{mu1,mu2,mu3} Finally, simultaneously negative magnetic permeability and electric permittivity in the optical telecommunicationfrequency range, and a hence negative refractive index, were experimentally demonstrated for metal-dielectric metamaterials \cite{NIMExp1,NIMExp2,NIMExp21}.
For review, see Ref. \cite{Sh}
Great progress was recently achieved in fabricating plasmonic NIMs, including chiral NIMs, and is described in Refs. \cite{Jas,SM,Zh,WeLi,SZh,Pl}

To satisfy the casuality principle, a NIM must be lossy \cite{caus1,caus2}. The majority of  NIMs realized to date contain  metal nanostructures. These structures introduce strong  losses inherent to metals that are difficult to avoid, especially in the visible range of frequencies.
Irrespective of their origin, losses constitute a major hurdle to the practical realization of the unique optical applications of these structures. Therefore, developing efficient loss-compensating techniques is of a paramount importance. So far, the most common approach to compensating losses in NIMs is associated with the possibility of embedding amplifying centers in the host matrix. The amplification is supposed to be provided through a population inversion between the energy levels of the embedded centers.
Such a method to remove absorption losses and dissipation by optical amplification, e.g. by introducing a lasing
medium instead of a passive dielectric into a negative-positive index composite and to utilize alternating amplifying layers instead of a bulk NIM, was suggested in \cite{amp1}. Although there are problems with this approach (for example, regarding amplified spontaneous noise, the influence of surface plasmons and the extreme localization of fields in optical amplification), it appears that it could furnish improved results \cite{amp2,amp3,amp4,amp5,Sh}. Raman amplification can be also used for removing the loss obstacle \cite{ampr}.
Exciting progress was achieved during the last year in producing population-inversion-based amplification in plasmonic metamaterials  \cite{Siv,Nog}, including recent a breakthrough with compensating losses in NIMs \cite{Ampl}.

The main emphasis in the studies of NIMs has been placed so far on
linear optical effects. Nonlinear optics in NIMs, especially frequency mixing, still remains a less-developed branch of photonics. It has been shown that NIMs, which include structural elements with non-symmetric current-voltage
characteristics -- its ``meta-atoms,'' can possess a nonlinear magnetic response at high frequencies \cite{Lap1,Zhar,Lap2,Gorkopa,Shadr}. Such meta-atoms have highly controllable magnetic and electric responses. Hence, NIMs can combine unprecedented linear and nonlinear electromagnetic properties.
The feasibility of crafting NIMs with nonlinear optical (NLO) responses in the optical wavelength range, such as  second-harmonic generation (SHG) and third-harmonic generation in magnetic metamaterials, has been experimentally demonstrated ~\cite{shg1e,shg2e,Kl}. On a fundamental level, the NLO response of nanostructured metamaterials is not yet completely understood or characterized. Nevertheless, it is well established that local-field-enhanced nonlinearities can be attributed to plasmonic nanostructures. Most negative-index (NI) plasmonic metamaterials are metal-dielectric composites made of metallic nanostructures shaped as split rings, horseshoes, fishnets, or chiral structures that are plunged into a dielectric host. This makes \emph{ab initio} calculations of nonlinear propagation processes a complex task. Such works fall into a special, rapidly expanding research category. Recent progress in studying NLO responses of plasmonic metamaterials, both in theory and proof-of-principle experiments, is described in Refs. \cite{Shad,Pet,Ze,Nie} and references therein.

The important properties of SHG in NIMs in the constant-pump approximation were discussed in Ref.\cite{Agr} for loss-less, semi-infinite materials and in Refs. \cite{Lens,Lens1} for a slab of a finite thickness. Unlike ordinary NLO materials, the latter appeared to bring major qualitative effects.
The possibility of exact phase-matching for waves with
counter-propagating energy-flows has been shown in Ref. \cite{KivSHG} for
the case when the fundamental wave falls in the negative-index
frequency domain and the second-harmonic (SH) wave lies in the positive-index domain.
The possibility of the existence of multistable nonlinear effects in
SHG was also predicted in Ref. \cite{KivSHG} As outlined here, absorption is one of the most challenging problems
that needs to be addressed for the realization of practical applications of NIMs. A
transfer of the near-field image into aSH frequency domain, where absorption may be lower, was proposed in REfs. \cite{Agr,Lens} as a possible means to overcome dissipative losses and thus enable the superlens. The propagation of
microwave radiation in nonlinear transmission lines, which are the
one-dimensional analog of NIMs, was investigated in Ref. \cite{Kozyr} The possibility to achieve conversion efficiency up to 12\% was shown in Ref.\cite{Sc} for phase-matched SHG in a strongly absorbing, bulk NIM. In that case, the input picosecond-pulse peak intensity was assumed to be about 500 MW/cm$^2$, the nonlinear susceptibility of the metamaterial was assumed to be $\chi^2 \sim$ 80 pm/V, and the attenuation depth of the NIM was about 50 $\mu$m. The possibility of achieving phase- and group-velocity matching was shown for the above-indicated pulses based on the Drude dispersion model for metals. Nonlinear-optical processes in NIMs possess unusual, sometimes counter-intuitive, properties. They are described below with the example of SHG in loss-less NIMs of a finite thickness taken from our recent paper \cite{SHG}. The extraordinary NLO properties of such propagation processes in NIMs, including three-wave mixing (TWM), four-wave mixing (FWM) and optical parametric amplification (OPA), also have been predicted and are in stark contrast with their counterparts in natural materials \cite{SHG,APB,met,OL,OLM,APL,LSh,APB09,OL09,SPI,JOA,WAS, EPJD,Ch}. Striking changes in the properties of nonlinear pulse propagation and temporal solitons \cite{Laz}, spatial solitons in systems with bistability \cite{Tas,Kos,Boa}, gap solitons \cite{Agu}, and optical bistability in layered structures including NIMs \cite{Lit,LitOL} have been revealed.  A review of some of the corresponding theoretical approaches is given in Refs. \cite{Gab,El} Frequency-degenerate multi-wave mixing and self-oscillations of counter-propagating waves in ordinary materials have been extensively studied earlier because of their easily achievable phase matching. Phase matching for TWM and four-wave mixing (FWM) of contra-propagating waves that are {far from degeneracy} seem impossible in ordinary materials and  presents a technical challenge in the metamaterials. It has become achievable only recently due to the advances in  nanotechnology \cite{Pas,Kh}. The possibility and extraordinary properties of TWM, including mirrorless self-oscillations, was proposed in Ref.~\cite{Har} (and references therein) more than 40 years ago (see also Refs. \cite{Vol,Yar}) for two co-propagating waves with nearly-degenerate frequencies that fall within an anomalous dispersion frequency domain and  gives rise to generation of a far-infrared difference-frequency counter-propagating wave in an anisotropic crystal. However, far-infrared radiation is typically strongly absorbed in crystals, which presents an unavoidable and strong detrimental factor. The proposal was not realized until recently. For the first time, a TWM backward-wave (BW) mirrorless optical parametrical oscillator (BWMOPO) with all three significantly different optical wavelengths was realized in Ref~\cite{Pas}. Phase-matching of counter-propagating waves has been achieved in a periodically poled NLO crystal with the period on a submicrometer scale. Both in the proposal \cite{Har} and in the experiment \cite{Pas}, the \emph{opposite} orientation of wavevectors was required to establish a distributed feedback and to produce mirrorless OPO. This was due to the fact that ordinary, positive-index crystals were proposed for the implementation. As outlined, a major technical problem in creating BWMOPO stems from the requirement of phase matching for the opposite orientation of wavevectors in PIMs. This paper reviews the counter-intuitive effects and unusual features in the energy exchange between coupled ordinary and contra-propagating backward electromagnetic waves associated with three types of nonlinear optical processes: second harmonic generation beyond the constant fundamental wave approximation, three-wave mixing, and four-wave mixing.

The extraordinary properties of second-harmonic generation in a frequency double-domain positive/negative matamaterial are studied in Section \ref{shg}.  Both semi-infinite and finite-length NIM slabs are considered and compared with each other and with ordinary PIMs for SHG.  The feasibility of a nonlinear-optical mirror converting the incident radiation into a reflected SH is shown and specific features of the process are investigated both for continuous-wave and pulsed regimes.  The paper is organized as follows. The ``backward'' features of the electromagnetic waves in NIMs are discussed in  Section \ref{pv}. The basic equations for the negative-index fundamental and positive-index SH waves are derived and the relevant Manley-Rowe relations are discussed  and compared with those in PIMs in Section \ref{bemr}.  The unusual spatial distributions of the field intensities for SHG in a NIM slab of a finite thickness are described in Section \ref{sl} for the model of a loss-free, double-domain metamaterial and a continuous-wave regime. These distributions are compared with their counterparts in ordinary PIMs (Section \ref{shgpim}) and semi-infinite NIM slabs (Section \ref{sis}). Section \ref{sis} also analyzes the properties of a nonlinear-optical mirror with a semi-infinite thickness that converts an incident fundamental beam into a reflected SH one. The role of absorption and phase mismatch is discussed in Section \ref{abs}. The extraordinary features associated with SHG in the pulsed regime, which include significant differences in the pulse shapes for the fundamental and SH radiations dependent on the intensity and on the pulse duration of the incident pulses, are considered in Section \ref{pul}.

The extraordinary features of coherent, NLO energy transfer from the control optical field(s) to the coupled contra-propagating negative-phase (negative index, NI) and positive-phase (positive index, PI) waves through three- and four-wave mixing and related optical parametric amplification are described in Sections \ref{twm} and \ref{fwm}. The uncommon phenomenon of generating a contra-propagating wave at an appreciably different difference frequency in the direction of reflection is  investigated. The feasibility and unparalleled features of such energy transfer are shown, which stem from the unusual fact that the energy flow of one of the coupled electromagnetic waves is \emph{contra-directed} relative to the others, whereas the wavevectors for all coupled waves remain \emph{parallel}. Such an opportunity makes phase matching of counter-propagating waves much easier and is offered by the backwardness of electromagnetic waves that is natural to NIMs. Consequently, distributed-feedback features become possible, while the antiparallel orientation of the wavevectors of the coupled waves is no longer required. The different properties attributed to such an all-optically tailored nonlinear optical mirror-amplifier for the case of a negative-index control field and for the alternative option of a negative-index idler are discussed and compared in Sections \ref{mcon} and \ref{mid}. The characteristic magnitudes of the required intensity of the control field and of the slab thickness are given in Section \ref{est1}.

The feasibility of independently crafting of a negative index and a resonantly enhanced four-wave mixing nonlinearity through embedded four-level centers is considered in Section \ref{fwm}. The characteristic values of the required density of the embedded resonant nonlinear-optical centers, their relaxation properties, and the intensity of the control field are given in Section \ref{est2}. The possibility of substituting a nanostructured negative-index metal-dielectric composites, which requires sophisticated fubrication techniques, with extensively used and studied ordinary crystals in order to simulate the unparalleled properties of coherent NLO energy exchange between the ordinary and backward waves is described in Section \ref{ph}. Finally, the possibilities of implementing originally strongly absorbing microscopic samples of plasmonic, nanostructured, metal-dielectric composites for remote, all-optical tailoring of the transparency and reflectivity of metamaterial films as well as the concepts of several options of unique, ultracompact photonic devices based on the outlined three- and four-wave processes are summarized in the concluding Section~\ref{con}.

\section{Second-harmonic generation in a lossy, dispersive double-domain slab}
\label{shg}
\subsection{Wavevectors and Poynting vectors for the fundamental and second-harmonic
waves in a double-domain metamaterial} \label{pv}
We consider a loss-free material that is left-handed at the fundamental frequency $\omega_1$ ($\epsilon_1<0$, $\mu_1<0$), whereas it is right-handed at the SH frequency $\omega_2=2\omega_1$
($\epsilon_2>0$, $\mu_2>0$). The relations between the vectors of the
electric, $\mathbf{E}$, and magnetic, $\mathbf{H}$, field components
and the wavevector $\mathbf{k}$ for an electromagnetic wave written as
\begin{eqnarray}
\mathbf{E}(\mathbf{r},t)&=&\mathbf{E}_0(\mathbf{r})\exp[-i(\omega t-\mathbf{%
k\cdot r})]+ c.c.,  \label{EM} \\
\mathbf{H}(\mathbf{r},t)&=&\mathbf{H}_0(\mathbf{r})\exp[-i(\omega t -\mathbf{%
k\cdot r})]+ c.c.,  \label{HM}
\end{eqnarray}
traveling in a loss-free medium with dielectric permittivity
$\epsilon$ and magnetic permeability $\mu$ are given by the equations
\begin{eqnarray}
&\mathbf{k}\times\mathbf{E} = ({\omega}/{c})\mu\mathbf{H},\quad\mathbf{k}%
\times\mathbf{H} =- ({\omega}/{c}) \epsilon\mathbf{E},&  \label{kh} \\
&\sqrt{\epsilon}{E}(\mathbf{r},t)=-\sqrt{\mu}{H}(\mathbf{r},t),&
\label{eh}
\end{eqnarray}
which follow from Maxwell's equations. Equations (\ref{kh}) show
that the vector triplet $\mathbf{E}$, $\mathbf{H}$ and $\mathbf{k}$
forms a right-handed system for the SH wave and a left-handed system
for the fundamental beam. Simultaneously negative material parameters ($\epsilon<0$ and
$\mu<0$) result in a negative refractive index $n= -
\sqrt{\mu\epsilon}$. As seen from Eqs. (\ref{EM}) and
(\ref{HM}), the phase velocity $\mathbf{v}_{ph}$ is co-directed with $%
\mathbf{k}$ and is given by $\mathbf{v}_{ph}=({\mathbf{k}}/{k})({\omega}/{%
k})=({\mathbf{k}}/{k})({c}/{|n|})$, where
${k}^{2}=n^{2}(\omega/{c})^{2}$. In contrast, the direction of the
energy flow (Poynting vector) $\mathbf{S}$ with respect to
$\mathbf{k}$ depends on the signs of $\epsilon $ and $\mu $:
\begin{equation}
\mathbf{S}(\mathbf{r},t) =\frac{c}{4\pi}[\mathbf{E}\times\mathbf{H}] =%
\frac{c^{2}}{4\pi\omega\epsilon}[\mathbf{H}\times\mathbf{k}\times\mathbf{H}]
=  \frac{c^{2}\mathbf{k}}{4\pi\omega\epsilon}H^{2} =\frac{c^{2}\mathbf{k}}{%
4\pi\omega\mu}E^{2}.  \label{s}
\end{equation}

Since we have taken the material to be loss-free, we assume here that all indices of $\epsilon$, $\mu$ and
$n$ are real numbers. Thus, the energy-flow $\mathbf{S}_1$ at
$\omega_1$ is directed opposite to $\mathbf{k}_1$, whereas
$\mathbf{S}_2$ is co-directed with $\mathbf{k}_2$.
\subsection{Basic equations and the Manley-Rowe relations for SHG process in NIMs and PIMs}\label{bemr}
\begin{figure}[h]
\begin{center}
\includegraphics[width=.6\columnwidth]{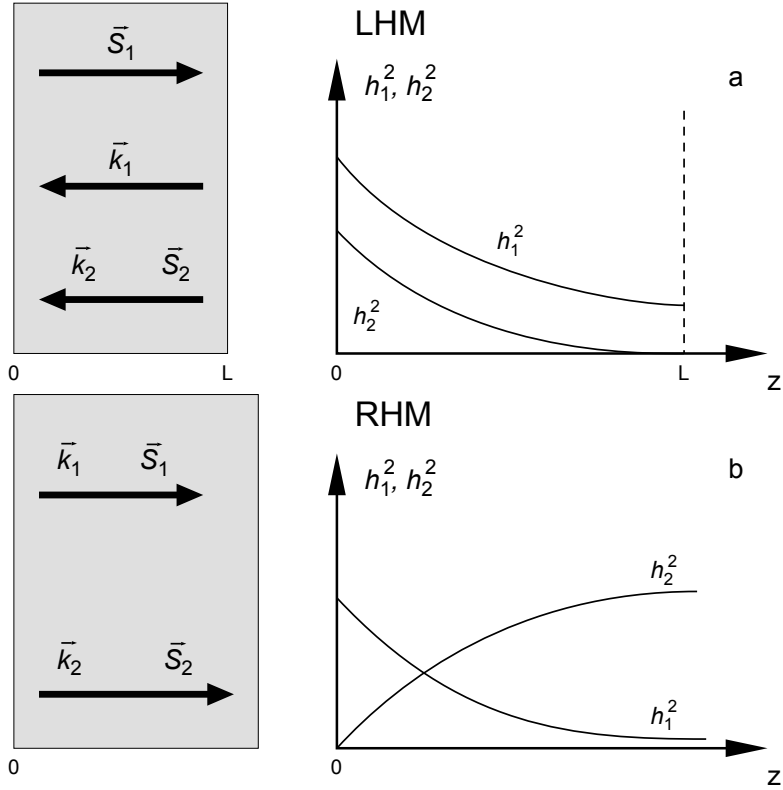}
\caption{\label{fig1} The difference in the phase-matching geometry and in the intensity distribution for the fundamental and the second-harmonic waves, $h_1^2$ and $h_2^2$, between a slab of the left-handed (negative-index) metamaterial, (a), and a right-handed material (ordinary, positive-index) material, (b). }
\end{center}
\end{figure}
\subsubsection{Basic equations}\label{be}
We assume that an incident flow of fundamental radiation $\mathbf{
S}_{1}$ at $\omega _{1}$ propagates along the z-axis, which is normal
to the surface of a metamaterial. According to (\ref{s}), the phase of
the wave at $\omega _{1}$ travels in the reverse direction inside the
NIM [Fig.\ref{fig1}(a)]. Because of the phase-matching requirement, the
generated SH radiation also travels backward with energy flow in the
same backward direction. This is in contrast with the standard
coupling geometry in a PIM [Fig.\ref{fig1}(b)].
As shown below, the basic features of the process associated with electric or magnetic nonlinearities are similar.  Following Ref. \cite{KivSHG}, we assume that a nonlinear
response is primarily associated with the magnetic component of the
waves. Then the equations for the coupled fields inside a NIM in the
approximation of slow-varying amplitudes acquire the form:
\begin{eqnarray}
{dA_{2}}/{dz} &=&i\sigma_2A_{1}^{2}\exp (-i\Delta kz),  \label{A2} \\
{dA_{1}}/{dz} &=&i\sigma_1A_{2}A_{1}^{\ast }\exp (i\Delta kz).
\label{A1}
\end{eqnarray}
Here, $\Delta k=k_{2}-2k_{1}$;\, $\sigma_1 =({\epsilon _{1}\omega
_{1}^{2}}/{k_{1}c^{2}})8\pi \chi ^{(2)}$,\, $\sigma_2=({\epsilon
_{2}\omega _{2}^{2}}/{k_{2}c^{2}})4\pi \chi^{(2)}$; $\chi^{(2)}$ is
the effective nonlinear susceptibility; $A_{2}$ and $A_{1}$ are the
slowly varying amplitudes of the waves with the phases traveling
against the z-axis,
\begin{equation}
{H}_{j}(z,t)={A}_{j}\exp [-i(\omega _{j}t+k_{j}z)]+c.c.,  \label{Az}
\end{equation}
$j=\{1,2\}$; $\omega _{2}=2\omega _{1}$; and $k_{1,2}>0$ are the
moduli of the wavevectors directed against the z-axis. We note that
according to Eq. (\ref{eh}), the corresponding equations for the
electric components can be written in a similar form, with $\epsilon
_{j}$ substituted by $\mu _{j}$ and vice versa. The factors $\mu _{j}$
were usually assumed to be equal to one in similar equations for PIMs.
However, this assumption does not hold for the case of NIMs, and this
fact dramatically changes many conventional electromagnetic relations.
\subsubsection{The Manley-Rowe relations}\label{mrr}
The Manley-Rowe relations \cite{MR56,MR} for the field intensities and for
the energy flows follow from Eqs. (\ref{s}) - (\ref{A1}):
\begin{equation}
\frac{k_{1}}{\epsilon
_{1}}\dfrac{d|A_{1}|^{2}}{dz}+\frac{k_{2}}{2\epsilon
_{2}}\dfrac{d|A_{2}|^{2}}{dz}=0,\quad
\dfrac{d|S_{1}|^{2}}{dz}-\dfrac{d|S_{2}|^{2}}{dz}=0. \label{MR1}
\end{equation}
The latter equation accounts for the difference in the signs of
$\epsilon _{1}$ and $\epsilon _{2}$, which brings radical changes to
the spatial dependence of the field intensities, as discussed below.

In order to outline the basic difference between the SHG process in
NIMs and PIMs, we assume in our further consideration that the phase-matching condition $k_{2}=2k_{1}$ is fulfilled. The
spatially-invariant form of the Manley-Rowe relations follows from
equation (\ref{MR1}):
\begin{equation}
|A_{1}|^{2}/\epsilon _{1}+|A_{2}|^{2}/\epsilon _{2}=C,  \label{I}
\end{equation}
where $C$ is an integration constant. With $\epsilon _{1}=-\epsilon
_{2}$ Equation (\ref{I}) takes the form:
\begin{equation}
|A_{1}|^{2}-|A_{2}|^{2}=C,  \label{D}
\end{equation}

Equations (\ref{I}) and (\ref{D}) predict a \emph{concurrent decrease} of the amplitudes of both waves along the z-axis so that the \emph{difference} between the squared amplitudes remains constant through the sample, as schematically depicted in Fig. \ref{fig1}(a). Hence, Eqs. ~(\ref{I}) and (\ref{D}) also predict that the \emph{difference} between the number of pairs of photons $\hbar\omega_1$ and the number of photons $\hbar\omega_2$ remains constant through the sample. This is in striking contrast with the requirement that the \emph{sum} of the squared amplitudes and the \emph{sum} of the corresponding photon numbers is constant in the analogous case in a PIM, as
schematically shown in Fig. \ref{fig1}(b) and described in textbooks on nonlinear optics.

Ultimately, the described extraordinary features of SHG in a double-domain NIM and their dependence on the slab geometry stem from the backwardness of electromagnetic waves in NIMs and the opposite directions of the energy flows in fundamental-harmonic (FH) and SH waves. As shown in the next sections for three-wave mixing processes, different depletion rates caused by different absorption indices at frequencies of the contra-propagating wave may qualitatively change their propagation properties, the distribution across the slab, and the output characteristics depending on the specific NLO propagation process.

We now introduce the real phases and amplitudes of the fields as
$A_{1,2}=h_{1,2}\exp (i\phi_{1,2})$. Then the equations for the real
amplitudes and phases
\begin{eqnarray}\label{psi}
&&dh_{2}/dz=\sigma_2h^2_{1}sin\Psi, \nonumber\\
&&dh_{1}/dz=-\sigma_1h_1h_2sin\Psi,\nonumber\\
&&d\Psi/dz=-(2\sigma_1h_2-\sigma_2h^2_1/h_2)cos\Psi,\\
&&\Psi\equiv\phi_2(z)-2\phi_1(z).\nonumber,
\end{eqnarray}
which follow from Eqs. (\ref{A2}) and (\ref{A1}), show that if any of
the fields becomes zero at any point, the integral (\ref{I})
corresponds to the solution with the constant phase difference
$2\phi_{1}-\phi_{2}=\pi /2$ over the entire sample.
\subsection{SHG in PIMs}\label{shgpim}
The equations for the slowly varying amplitudes corresponding to the
ordinary coupling scheme in a PIM, shown in  Fig. \ref{fig1}(b), are
readily obtained from Eqs. (\ref{A2}) - (\ref{Az}) by changing the
signs of $k_{1}$ and $k_{2}$. This does not change the integral
(\ref{I}); more importantly, the relation between $\epsilon_{1}$ and
$\epsilon_{2}$ required by phase matching now changes to
$\epsilon_{1}=\epsilon_{2}$, where both constants are positive. The
phase difference remains the same. Because of the boundary conditions
$h_{1}(0)=h_{10}$ and $h_{2}(0)=h_{20}=0$, the integration constant
becomes $C=h_{10}^{2}$. Thus, the equations for the real amplitudes in
the case of a PIM acquire the form:
\begin{eqnarray}
&&h_{1}(z)=\sqrt{h_{10}^{2}-h_{2}(z)^{2}},  \label{D3} \\
&&dh_{2}/{dz}=\kappa \lbrack h_{10}^{2}-h_{2}(z)^{2}],  \label{h24}
\end{eqnarray}
with the known solution
\begin{eqnarray}
h_{2}(z) &=&h_{10}\tanh (z/z_{0}),  \label{rhm2} \\
h_{1}(z) &=&h_{10}/\cosh (z/z_{0}),\,z_{0}=[\kappa h_{10}]^{-1}.
\label{rhm1}
\end{eqnarray}
Here, $\kappa =({\epsilon_{2}\omega_{2}^{2}}/{k_{2}c^{2}})4\pi
\chi_{eff}^{(2)}$. \emph{The solution has the same form for an
arbitrary slab thickness} with decreasing fundamental and increasing
SH squared amplitudes along the z-axis, as shown schematically in  Fig.
\ref{fig1}(b).
\subsection{SHG in a NIM slab}
\label{sl} 
\begin{figure}[h]
\begin{center}
\includegraphics[width=.6\textwidth]{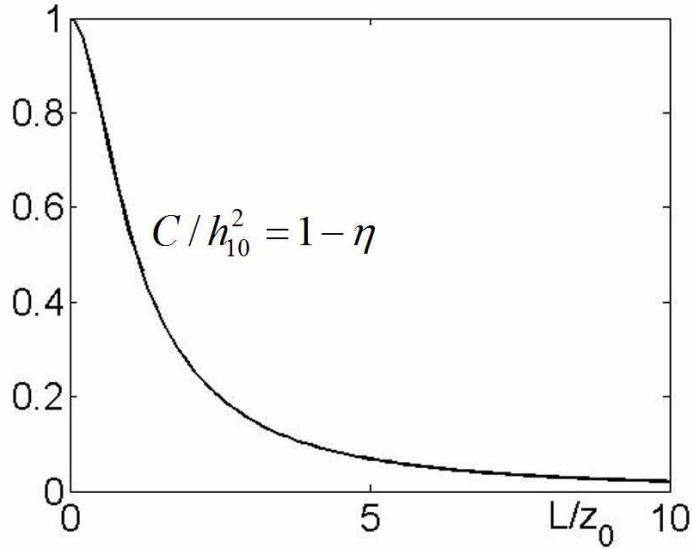}
\caption{The normalized integration constant $C/h_{10}^{2}$ and the
energy conversion efficiency $\protect\eta $ vs. the normalized length
of a NIM slab.} \label{f2}
\end{center}
\end{figure}

\begin{figure}[h]
\begin{center}
\includegraphics[height=.6\textwidth]{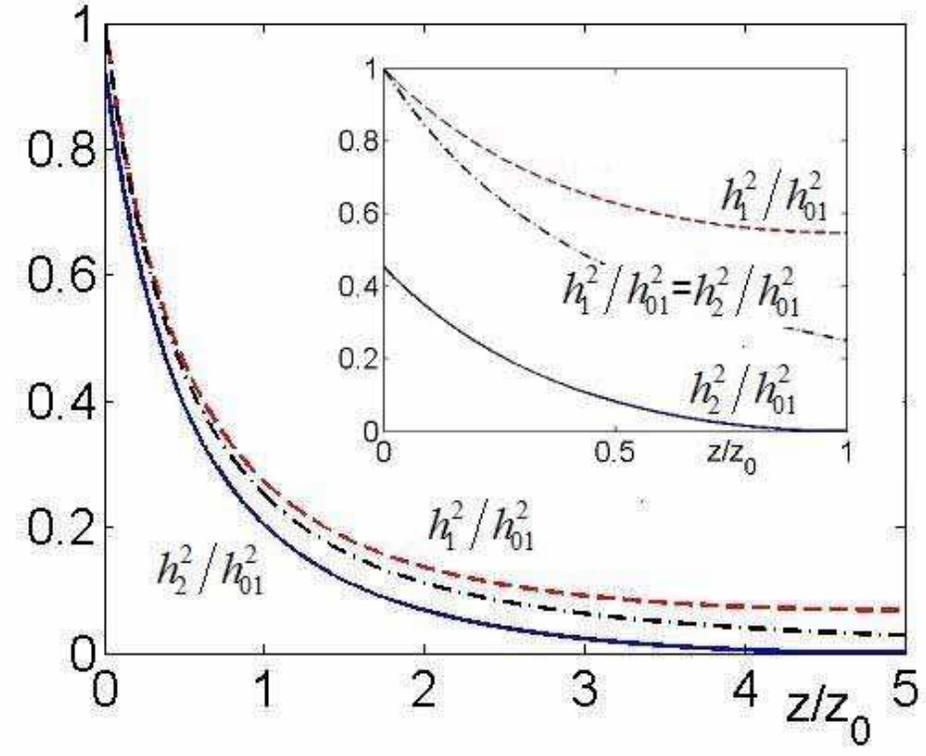}
\caption{The squared amplitudes for the fundamental wave (dashed
line) and SHG (solid line) in a loss-less NIM slab of a finite
length. Inset: the slab has a length equal to one conversion length.
Main plot: the slab has a length equal to five conversion lengths. The
dash-dot lines show the energy conversion for a semi-infinite NIM.}
\label{f3}
\end{center}
\end{figure}

Now consider phase-matched SHG in a loss-less NIM slab of a
finite length L. Equations (\ref{A2}) and (\ref{D}) take the form:
\begin{eqnarray}
h_{1}(z)^{2} &=&C+h_{2}(z)^{2},  \label{D1} \\
dh_{2}/{dz} &=&-\kappa \lbrack C+h_{2}(z)^{2}].  \label{h12}
\end{eqnarray}
Taking into account the \emph{different boundary conditions in a NIM
as compared to a PIM}, $h_{1}(0)=h_{10}$ and $h_{2}(L)=0$, the
solution to these equations is
\begin{eqnarray}
h_{2} &=&\sqrt{C}\tan [\sqrt{C}\kappa (L-z)],  \label{h22} \\
h_{1} &=&\sqrt{C}/\cos [\sqrt{C}\kappa (L-z)],  \label{h11}
\end{eqnarray}
where the integration parameter $C$ \emph{depends on the slab
thickness $L$} and on the amplitude of the incident fundamental
radiation as
\begin{equation}
\sqrt{C}\kappa L=\cos ^{-1}(\sqrt{C}/h_{10}).  \label{C}
\end{equation}
Thus, the spatially invariant field intensity difference between
the fundamental and SH waves in NIMs depends on the slab thickness,
which is in \emph{strict contrast} with the case in PIMs. As seen from
Equation (\ref{D1}), the integration parameter
$C=h_{1}(z)^{2}-h_{2}(z)^{2}$ now represents the deviation of the
conversion efficiency $\eta =h_{20}^{2}/h_{10}^{2}$ from unity:
$(C/h_{10}^{2})=1-\eta $. Figure \ref{f2} shows the dependence of this
parameter on the conversion length $z_{0}=(\kappa h_{10})^{-1}$.
The figure shows that for a conversion length of 2.5, the NIM slab,
which acts as nonlinear mirror, provides about 80\% conversion of the
fundamental beam into a reflected SH wave. Figure \ref{f3} depicts the
field distribution along the slab length. One can see from the figure, with an increase in slab length (or intensity of the fundamental wave), the gap between the two plots decreases while the conversion
efficiency increases (comparing the main plot and the inset).
\subsection{SHG in a semi-infinite NIM}
\label{sis} Now we consider the case of a semi-infinite NIM at $z>0$.
Since both waves disappear at $z\rightarrow \infty $ due to the entire
conversion of the fundamental beam into SH, $C=0$. Then equations
(\ref{D1}) and (\ref {h12}) for the amplitudes take the simple form
\begin{eqnarray}
&&h_{2}(z)=h_{1}(z),  \label{D2} \\
&&dh_{2}/{dz}=-\kappa h_{2}^{2}.  \label{h23}
\end{eqnarray}
Equation (\ref{D2}) indicates 100\% conversion of the incident
fundamental wave into the reflected second harmonic at $z=0$ in a
loss-less, semi-infinite medium provided that the phase-matching
condition $\Delta k=0$ is fulfilled. The integration of (\ref{h23})
with the boundary condition $h_{1}(0)=h_{10}$ yields
\begin{equation}
h_{2}(z)=\dfrac{h_{10}}{(z/z_{0})+1},\,z_{0}=(\kappa h_{10})^{-1}.
\label{si}
\end{equation}
Equation (\ref{si}) describes a \emph{concurrent decrease} of both
waves of \emph {equal amplitudes} along the z-axis; this is shown by the
dash-dot plots in Fig. \ref{f3}. For $z\gg z_{0}$, the dependence is
inversely proportional to $z$. These spatial dependencies, shown in Fig. \ref{f3}, are also in striking contrast with those for the conventional process of SHG in a PIM, which are shown in various
textbooks [compare, for example, with Fig.\ref{fig1}(b)].

\subsection{SHG  in a lossy dispersive NIM slab}\label{abs}
It is convenient to introduce effective  amplitudes, $a_{j}$, and
nonlinear coupling parameters, $X_{j}$, which
are defined as
\begin{eqnarray*}
a_{j}=\sqrt{|\mu_j|/k_j}A_j,\quad
X=\sqrt{k_1^2k_2/|\mu_1^2\mu_2|}  4\pi\chi^{(2)}_{\rm eff}.
\label{ec}
\end{eqnarray*}
Here, $\chi^{(2)}_{\rm eff}$ is the effective nonlinear susceptibility, and the quantities  $|a_j|^2$ are proportional to the photon numbers in the energy fluxes.

In the general case of SHG in a dispersive double-domain material with $\alpha(\omega_1)\neq\alpha(\omega_2)$ and $\Delta k=k_{2}-2k_{1}~\neq0$, the equations for the amplitudes $a_j$ take the form:
\begin{eqnarray}
&&da_1/dz=-i2Xa^*_1a_2\exp(i\Delta k z)-(\alpha_1/2)a_1,\label{am1}\\
&&da_2/dz=iXa_1^2\exp(-i\Delta k z)+(\alpha_2/2)a_2.\label{am2}
\end{eqnarray}
Here,  $\alpha_{1,2}$ are the absorption indices at the corresponding frequencies. The equations account for opposite signs of $\epsilon_1$, $\mu_1$ and $\epsilon_2$, $\mu_2$. Three differences distinguish these equations from their counterparts in ordinary PIMs -- opposite signs with the NLO coupling factors, opposite signs with the absorption indices, and boundary conditions to be applied to the opposite sides of the slab. These differences stem from the backwardness of the fundamental wave in the NIM in this case. Note that, for a NIM with an electric nonlinearity, the corresponding equations for the electric components can be written in a similar form, with $\epsilon
_{j}$ substituted by $\mu _{j}$ and vice versa.

\begin{figure}[h!]
\begin{center}
\resizebox{.49\columnwidth}{!}{
\includegraphics{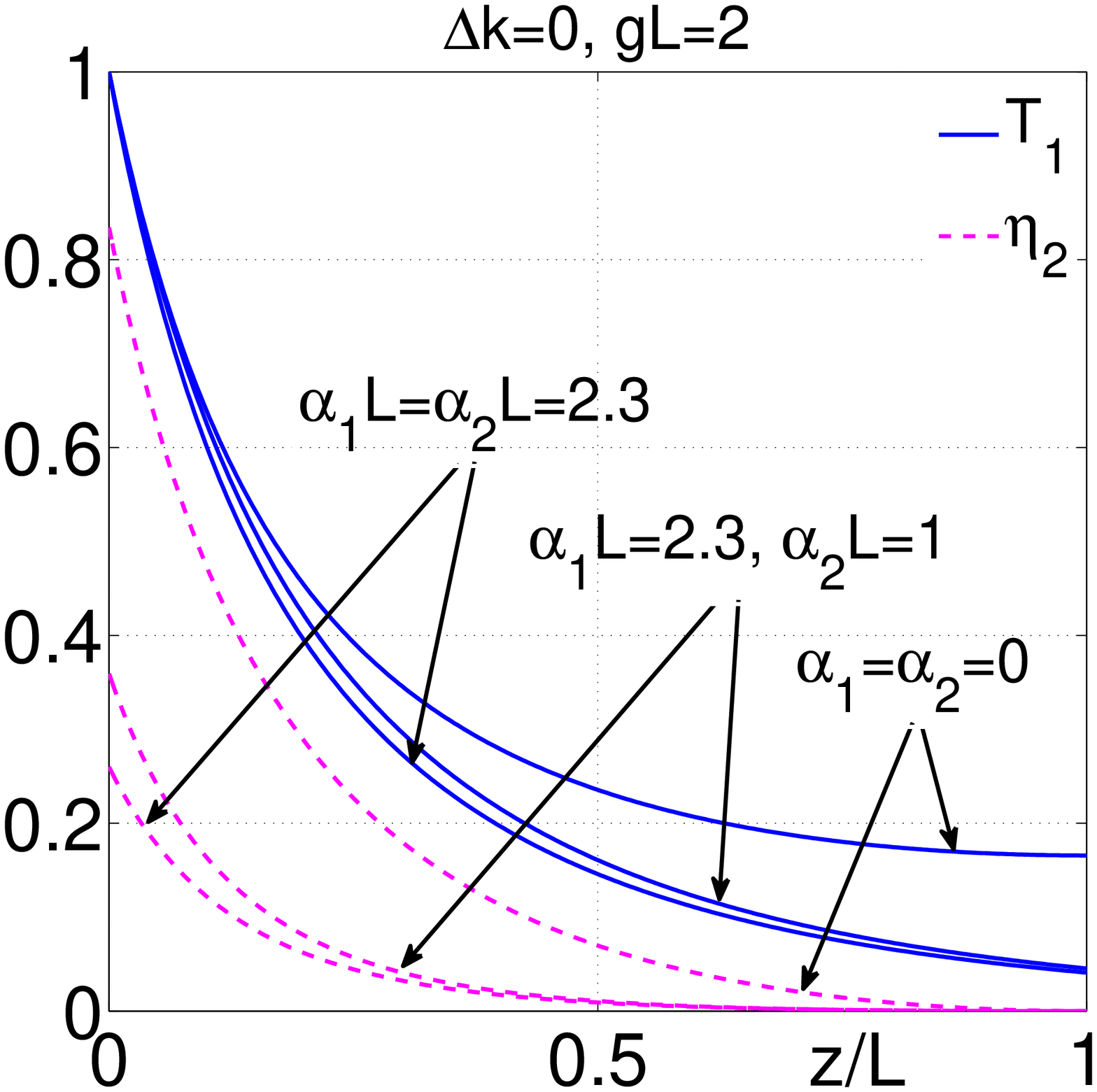}}
\resizebox{.49\columnwidth}{!}{
\includegraphics{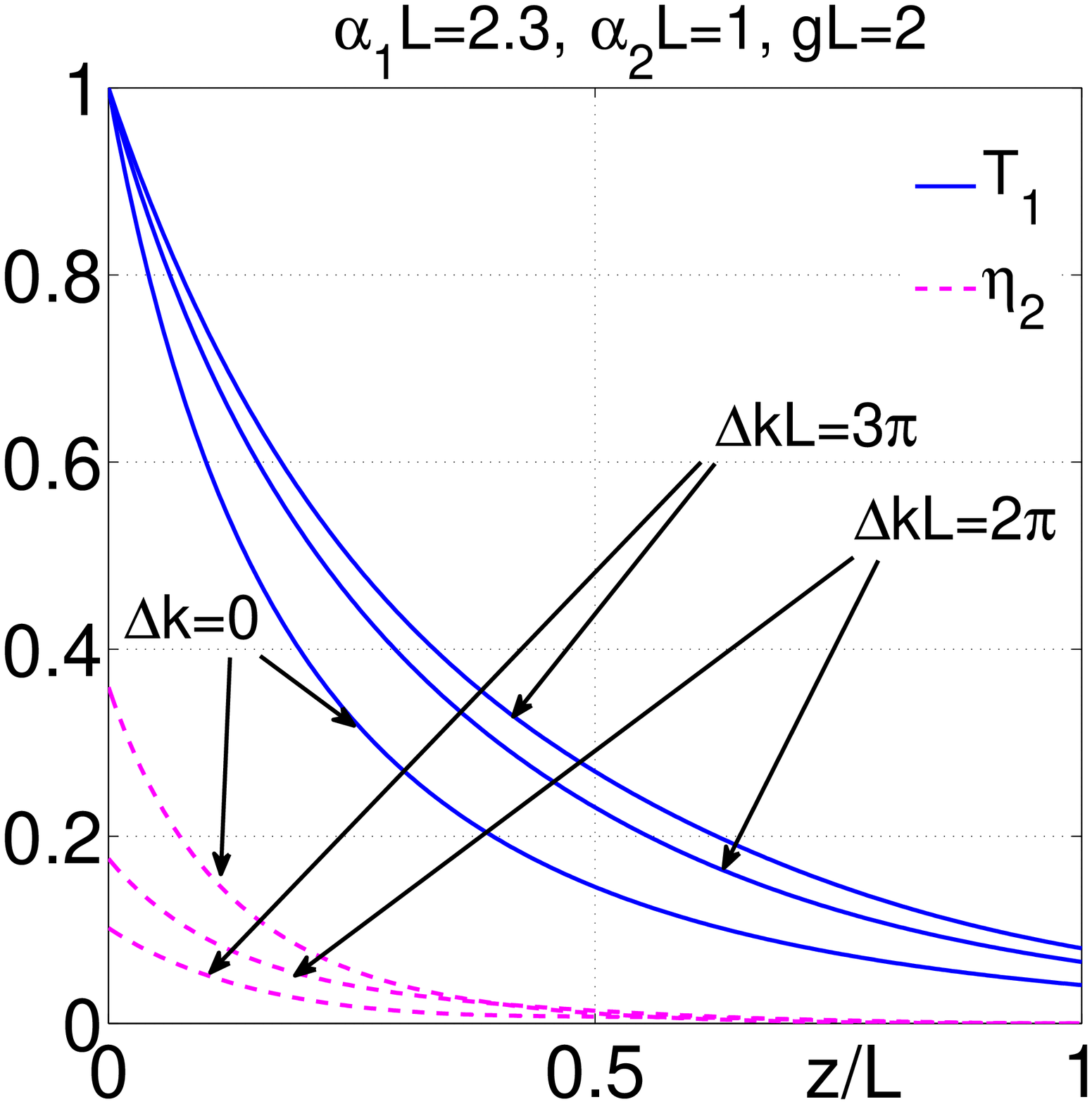}}\\
\hspace{5mm}(a)\hspace{80mm}(b)
\end{center}
\caption{\label{fig2} Effect of absorption (a) and phase mismatch (b) on the distribution of the FH and SH fields across the slab and on the output values. }
\end{figure}
\begin{figure}[h!]
\begin{center}
\resizebox{.49\columnwidth}{!}{
\includegraphics{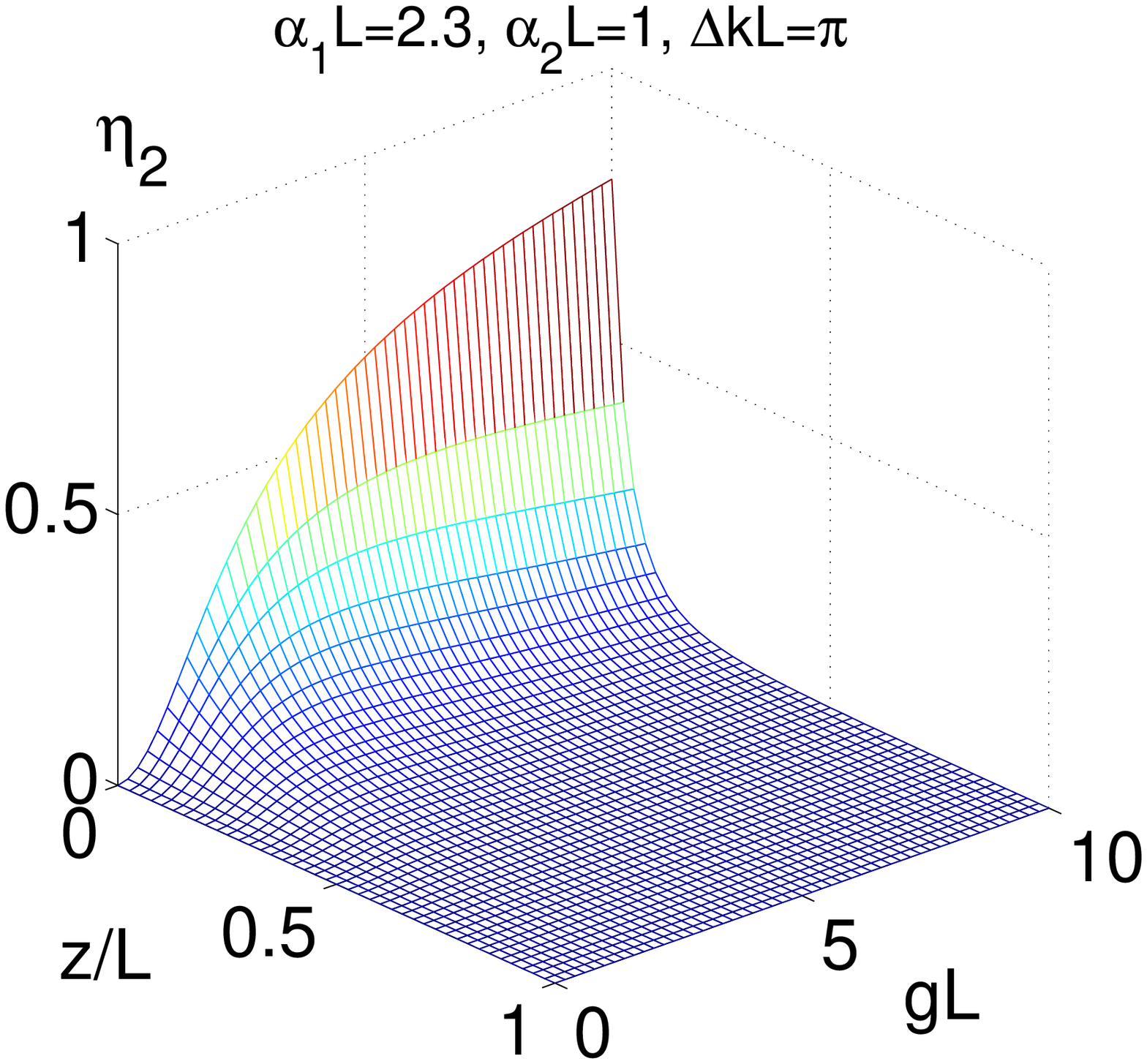}}
\resizebox{.489\columnwidth}{!}{
\includegraphics{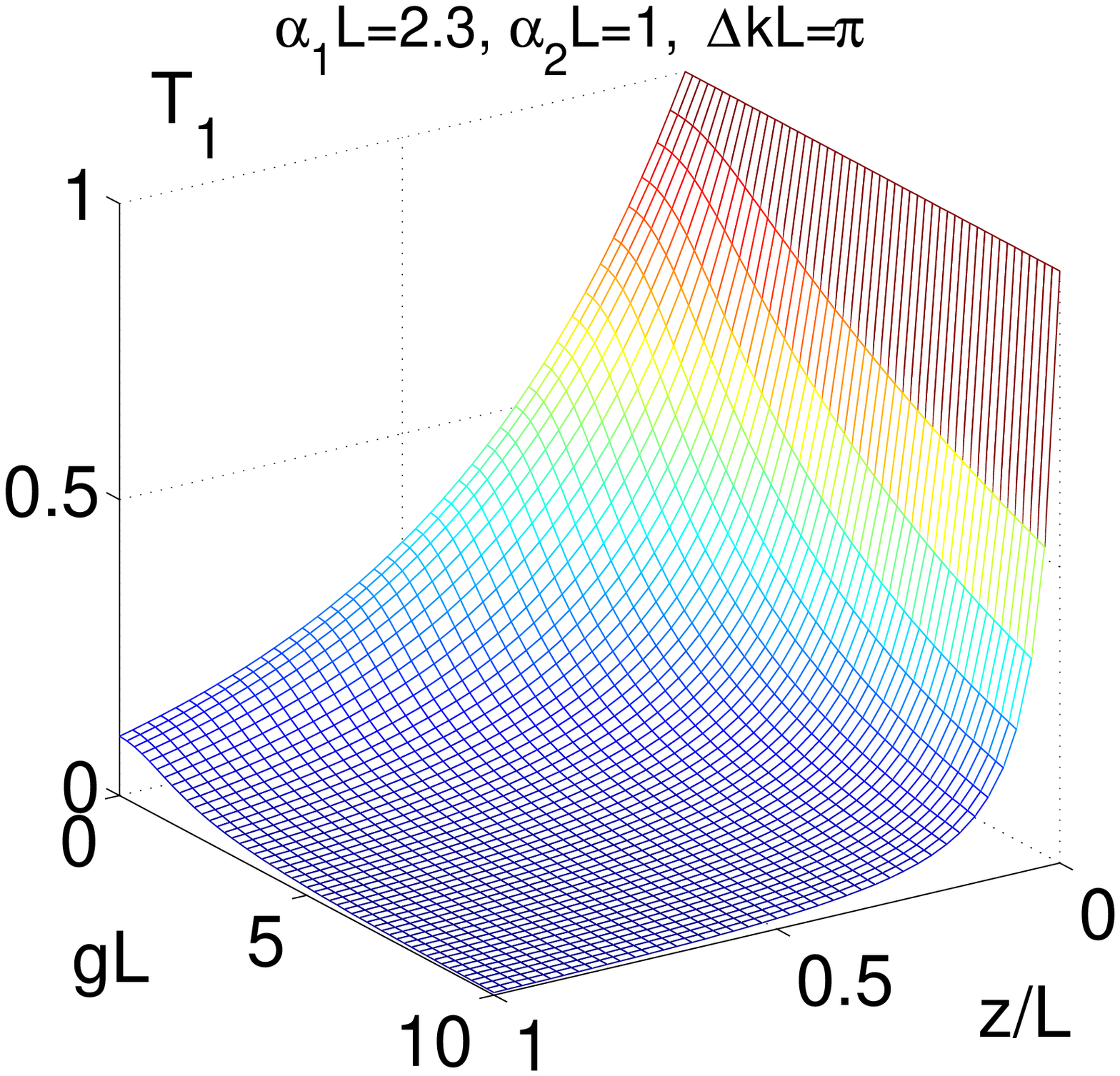}}\\
(a)\hspace{80mm}(b)
\end{center}
\caption{\label{fig3}  Dependence of the distribution of the FH and SH fields across the slab and of their output values on the intensity of the input fundamental beam for  $\Delta kL=\pi$.}
\end{figure}

Equations (\ref{am1})-(\ref{am2}) do not have analytical solutions. Figures~\ref{fig2} and \ref{fig3} depict the results of a numerical analysis of the chief features of SHG in a dispersive, absorptive slab for boundary conditions  $a_1(z=0)=a_{10}$, $a_2(z=L)=0$. The quantity $T_1=|a_1(z)|^2/|a_{10}|^2$ represents the distribution of the fundamental beam inside the slab determined by both absorption and energy conversion as well as the ultimate slab transparency,  $T_1(z=L)$. The quantity $\eta_2=2|a_2(z)|^2/|a_{10}|^2$ represents the quantum conversion efficiency of the process, $g=Xa_{10}$, and $g^{-1}$ is the nonlinear conversion length, i.e., the characteristic slab thickness that is required for significant energy conversion and is dependent on the input intensity of the fundamental field.
The plot for  $\alpha_{1}=\alpha_{2}=0$ in Fig.~\ref{fig2}(a) displays two curves with a constant gap, which determines the output value of about 80\% for the reflected beam ($z=0$) at $2\omega_1$ for the given input intensity of the fundamental field at $\omega_1$. By comparing this plot with the others, we see that absorption causes variable a gap that increases in the output area $z\geq0$, so that the output value of SHG drops to about 40\% for $\alpha_{1}L=2.3$ and $\alpha_{2}L=1$ and the same intensity of the input fundamental beam. For $\alpha_{1}L=2.3$ and $a_{10}\rightarrow0$, the transparency of the slab at $\omega_1$ reaches up to 10\%, which is characteristic for plasmonic samples. Figure \ref{fig2}(b) shows that the conversion efficiency decreases with an increase in phase mismatch, however it is less than it would be for the ordinary coupling schemes in PIMs.
Figure~\ref{fig3}(a) shows that, for the given parameters, the SH field concentrates in the area close to the exit facet of the slab at $z=0$. The dependence on $g$ at $z=0$ indicates that a significant part of the incident beam can be converted into SH for $gL\sim1$ even for the given phase mismatch and absorption rates. This is shown in Fig.~\ref{fig3}(b), which depicts the significant depletion of the FH field with an increase of its input intensity and the slab length (parameter $gL$).

\subsection{Pulsed SHG in a transparent NIM slab}\label{pul}
The equations for the quantities $a_1$ and $a_2$ introduced above are
\begin{eqnarray}\label{eq1}
  \frac1{v_1}\frac{\p a_1}{\p t}+ \frac{\p a_1}{\p z}&=& -i2ga_1^*a_2\exp{(i\Delta kz)}-\frac{\alpha_1}2a_1 \nonumber\\
  -\frac1{v_2}\frac{\p a_2}{\p t}+ \frac{\p a_2}{\p z}&=& iga_1^2\exp{(-i\Delta kz)}+\frac{\alpha_2}2a_2.
\end{eqnarray}
Here, $|a_{1,2}|^2$ are the above-introduced slowly varying amplitudes that are proportional to the photon numbers in the energy fluxes corresponding to the pulse maximum, and $v_i$ are the group velocities for the corresponding pulses.
After introducing  $\xi=z/L$ and $\tau=t/\Delta\tau$, $d=L/v_1\Delta\tau$, where  $d$  is the slub thickness $L$ reduced by the input pulse length and $\Delta\tau$ is the duration of the input fundamental pulse, Eqs.~(\ref{eq1}) take the form:
\begin{eqnarray}\label{eq2}
  d\frac{\p a_1}{\p \tau}+ \frac{\p a_1}{\p \xi}&=& -i2\tilde{g}a_1^*a_2\exp{(i\Delta \tilde{k}\xi)}-\frac{\tilde{\alpha}_1}2a_1 \nonumber\\
  -d\frac{\p a_2}{\p \tau}+ \frac{\p a_2}{\p \xi}&=& i\tilde{g}a_1^2\exp{(-i\Delta \tilde{k}\xi)}+\frac{\tilde{\alpha}_2}2a_1
\end{eqnarray}
The input pulse shape is taken as being close to a rectangular form
\begin{equation}
F(\tau)=0.5\left(\tanh\frac{\tau_0+1-\tau}{\delta\tau}-\tanh\frac{\tau_0-\tau}{\delta\tau}\right),
\end{equation}
where $\delta\tau$ is the duration of the pulse front and tail, and $\tau_0$ is the shift of the front relative to $t=0$. The magnitudes $\delta\tau=0.01$ and $\tau_0=0.5$ have been selected for numerical simulations.

\begin{figure}[ht]
\begin{center}
\includegraphics[width=.48\textwidth]{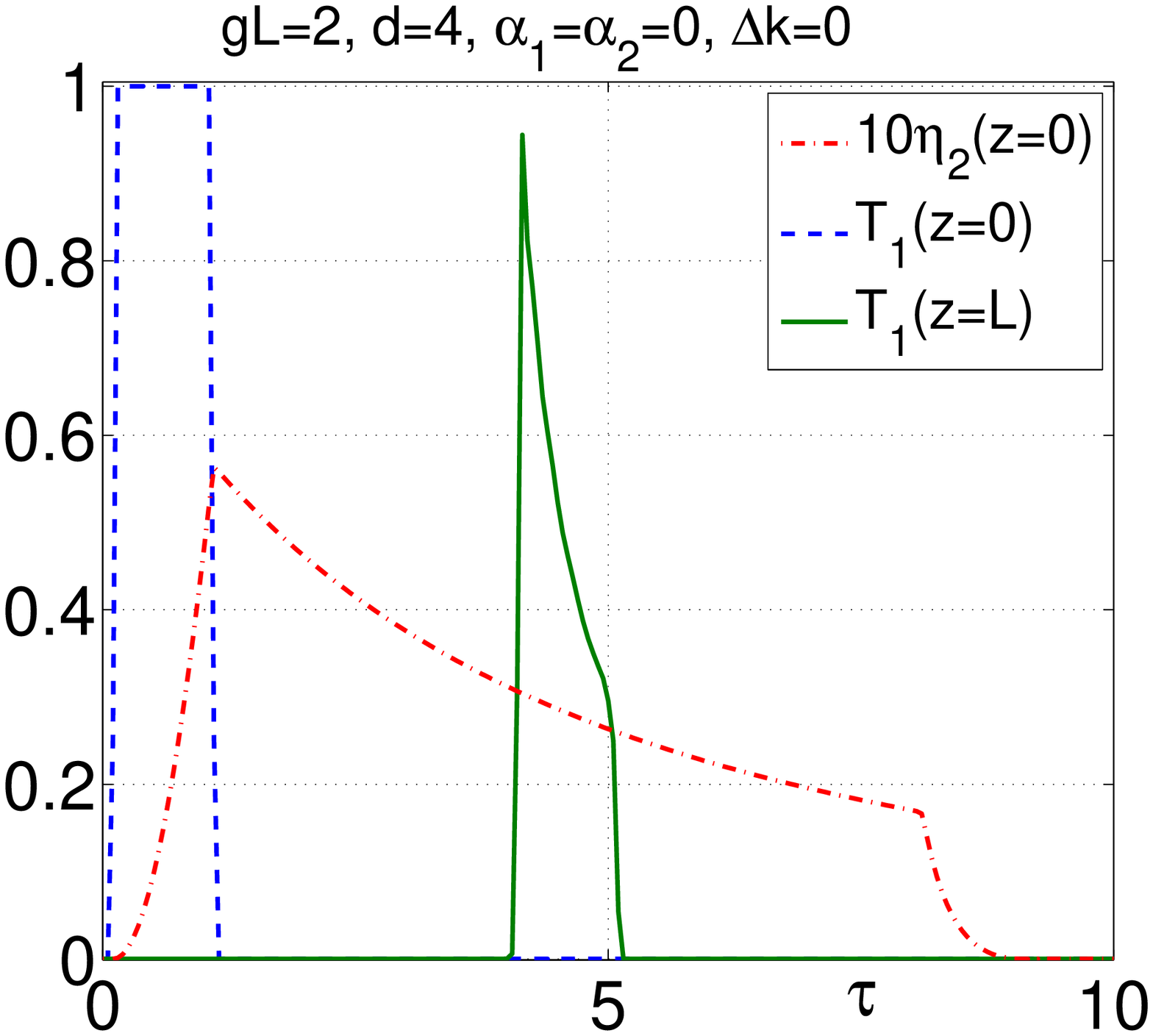}
\includegraphics[width=.48\textwidth]{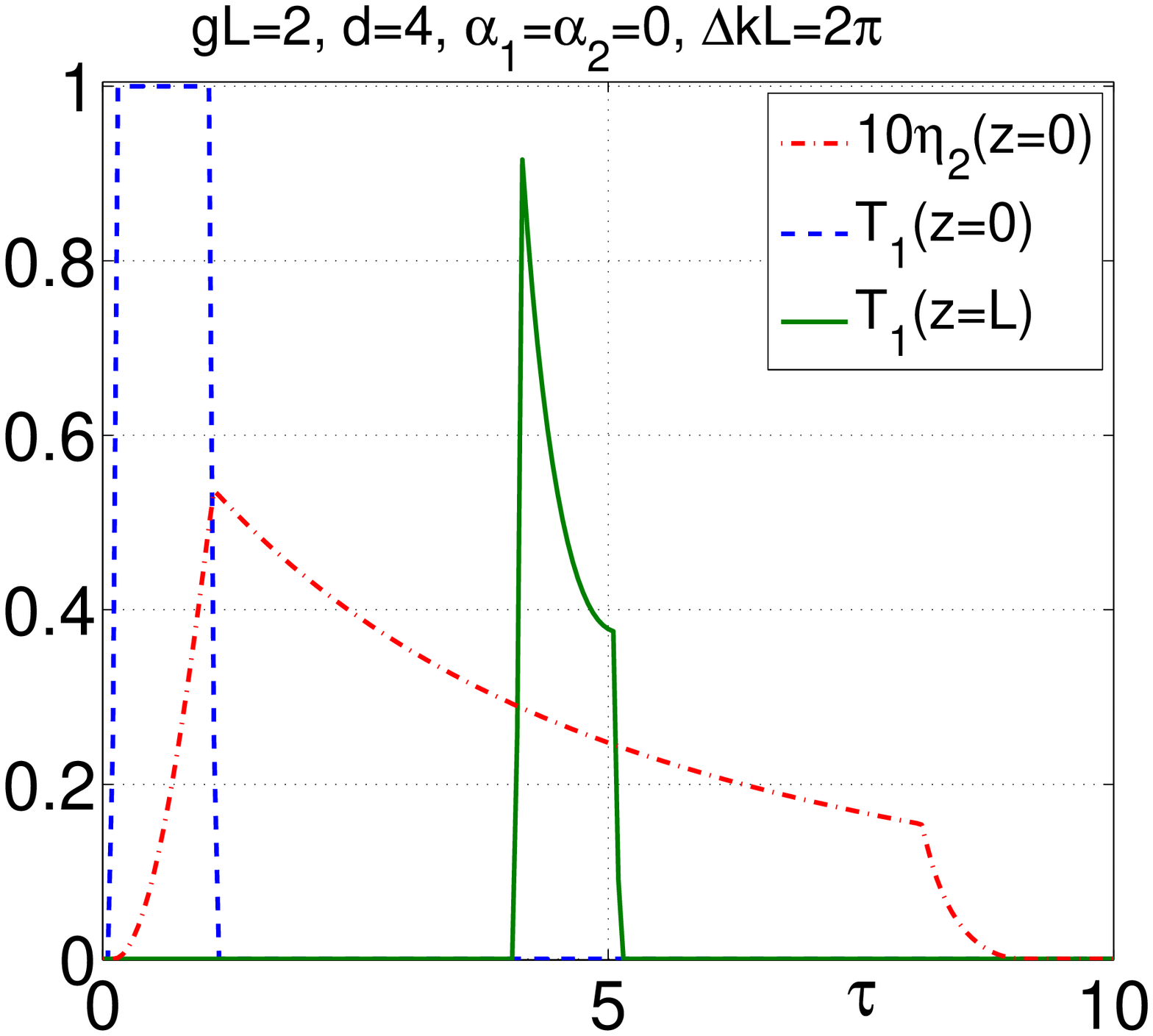}\\
(a)\hspace{80mm} (b)\\
\includegraphics[width=.48\textwidth]{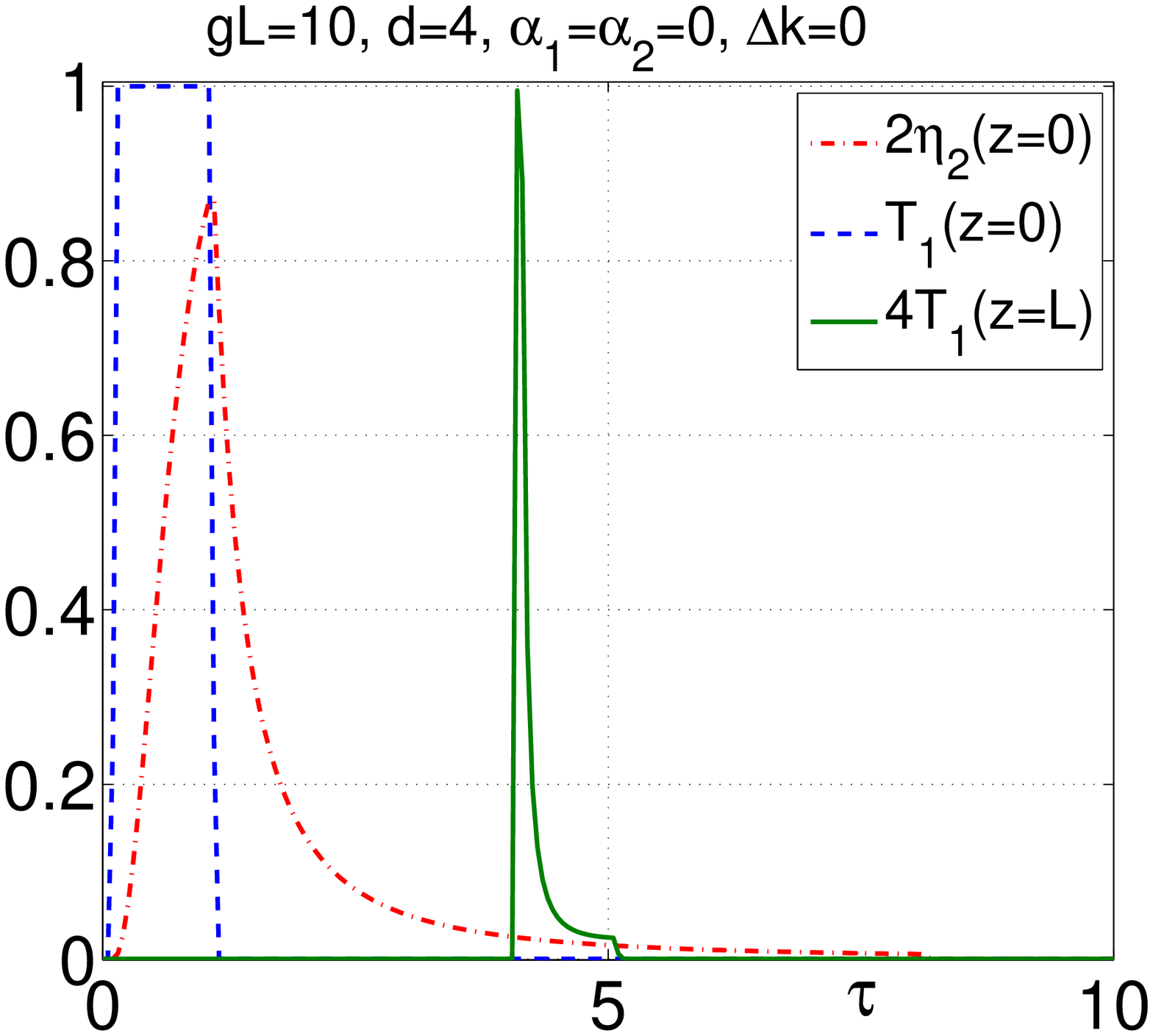}
\includegraphics[width=.48\textwidth]{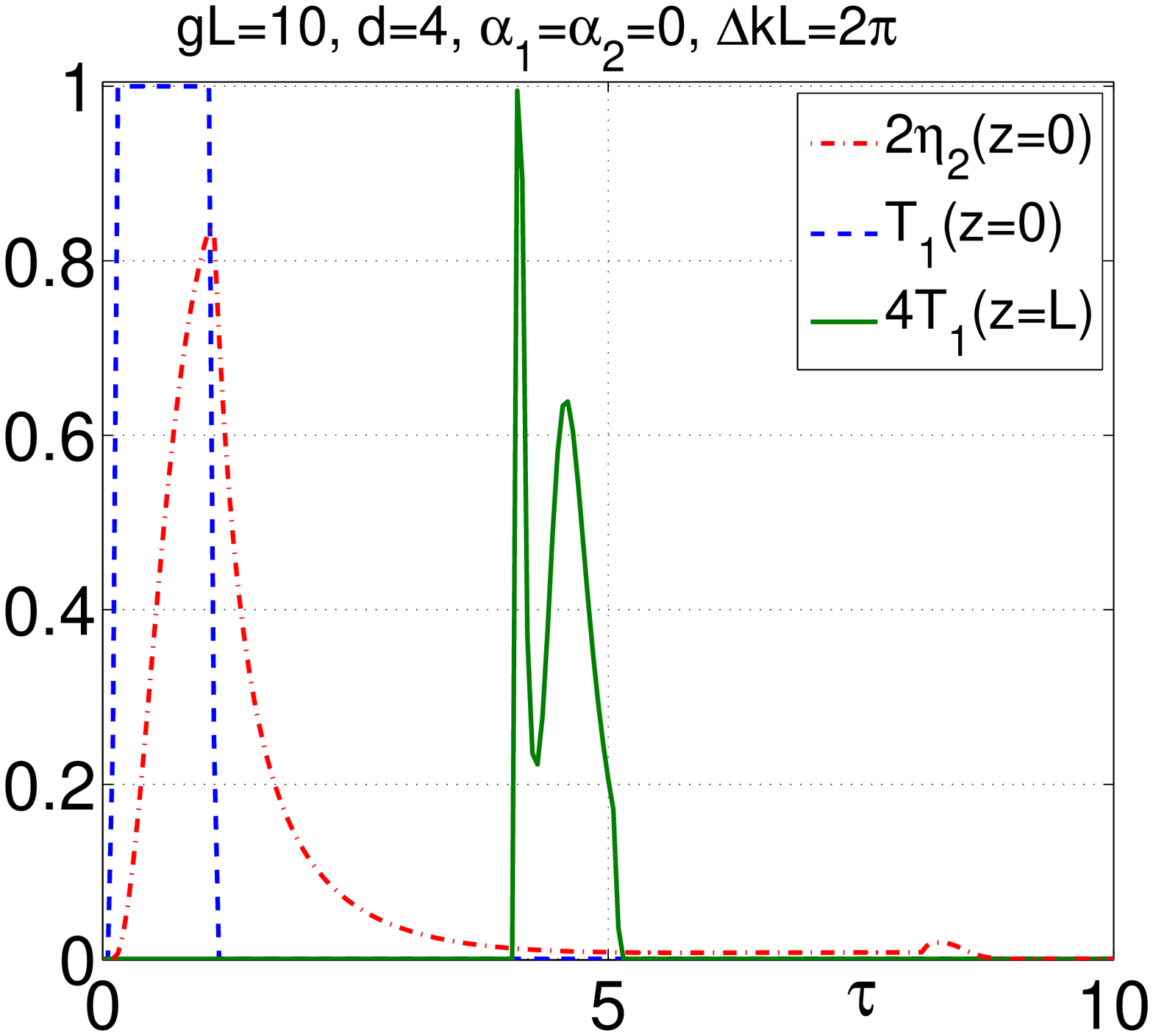}\\
(c)\hspace{80mm} (d)\\
\end{center}
\caption{\label{fi1} Input $T_1(z=0)$  and output $T_1(z=L)$  pulses of fundamental and negative-index SH radiation  $\eta_2(z=0)$  for relatively short pulses of fundamental radiation. (a) Input pulse area $S_{10}=0.9750$; output pulse areas $S_{1L}= 0.5031$, $S_{20}= 0.2392$. (c)  Input pulse area $S_{10}=0.9750$; output pulse areas $S_{1L}= 0.0396$, $S_{20}= 0.4742$.}
\end{figure}
\begin{figure}[htbp]
\begin{center}
\includegraphics[width=.48\textwidth]{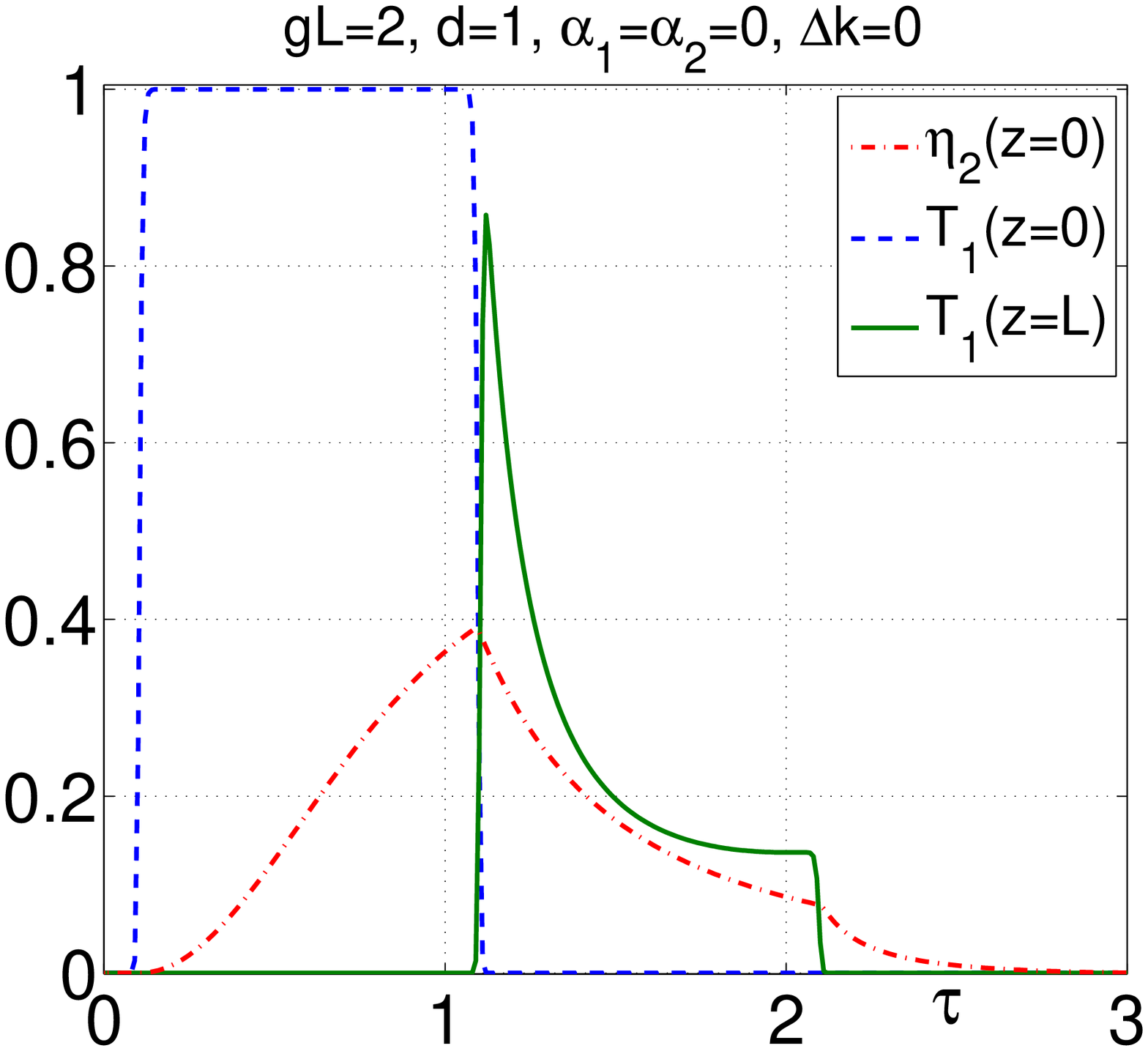}
\includegraphics[width=.48\textwidth]{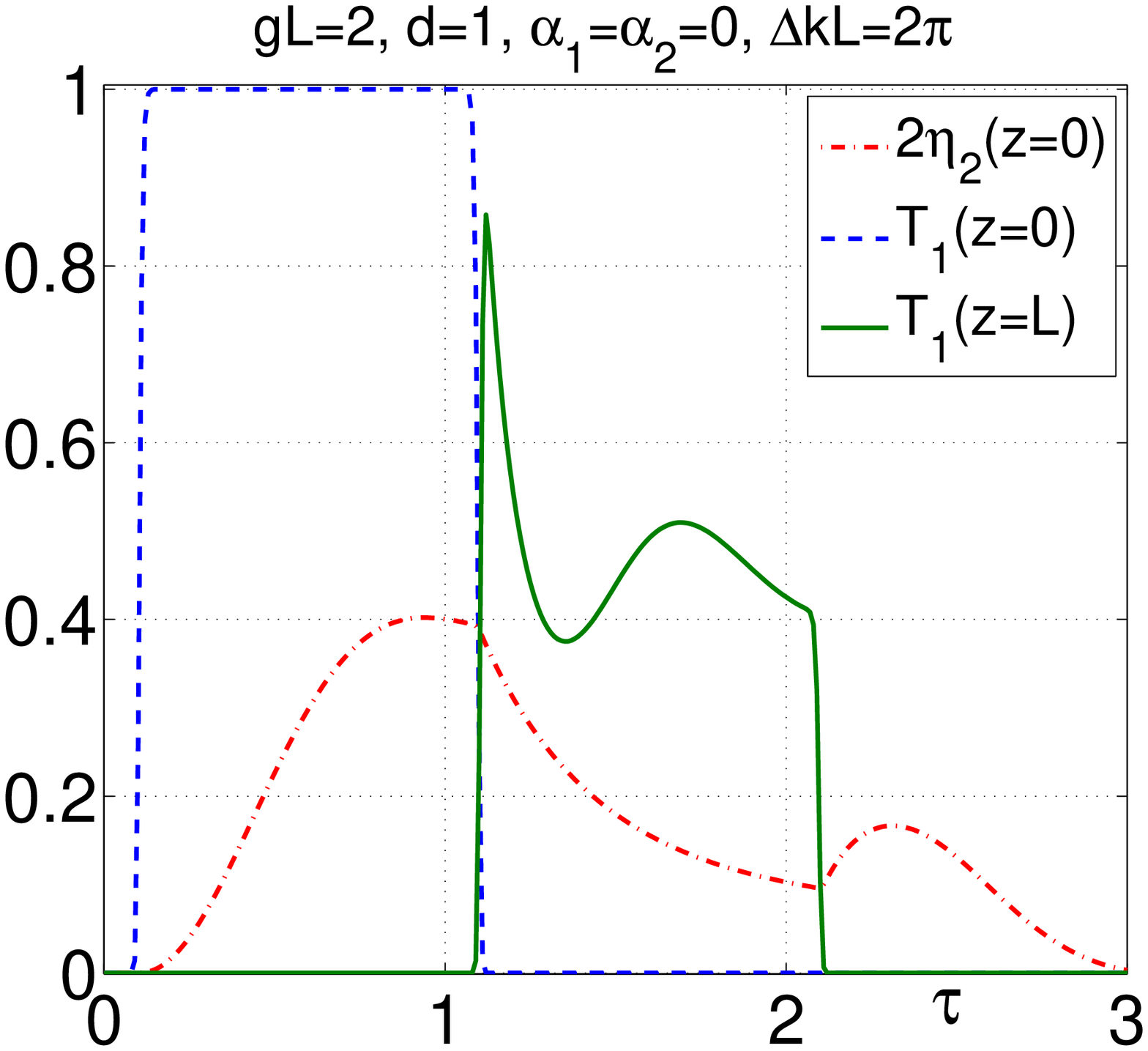}\\
(a)\hspace{80mm} (b)\\
\includegraphics[width=.48\textwidth]{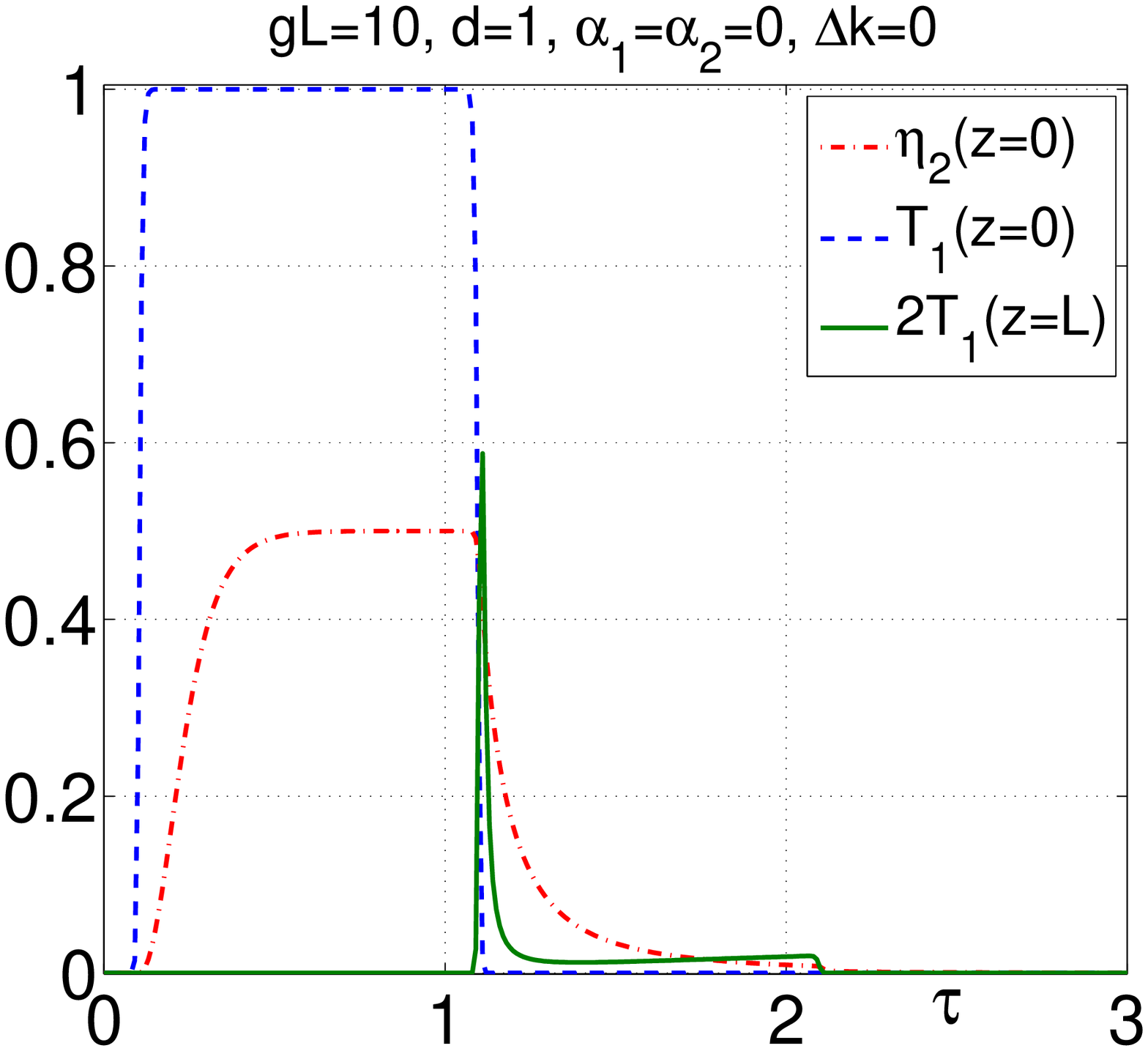}
\includegraphics[width=.48\textwidth]{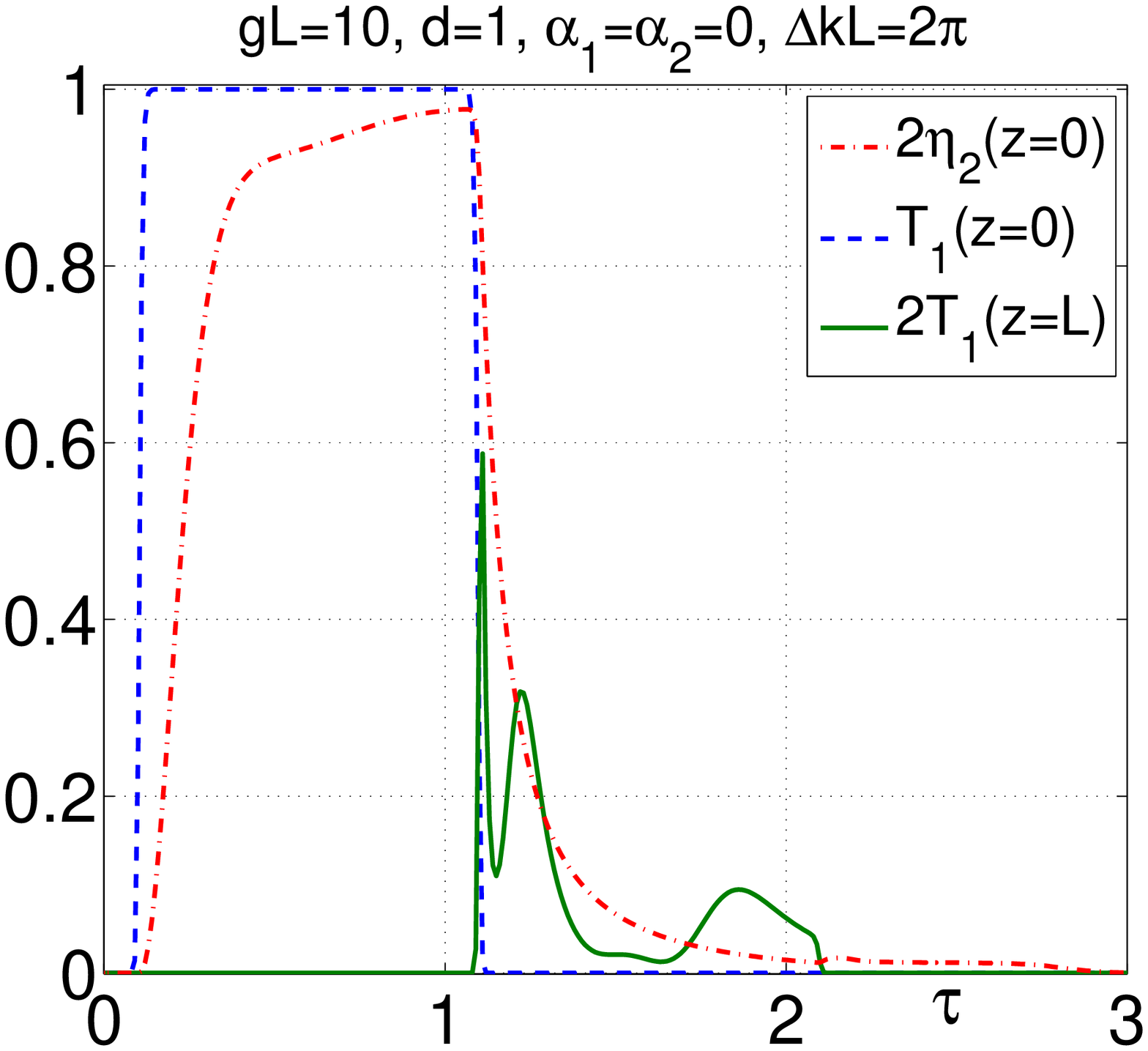}\\
(c)\hspace{80mm} (d)\\
\end{center}
\caption{\label{fi5}  Input $T_1(z=0)$  and output  $T_1(z=L)$  pulses of fundamental and negative-index SH radiation $\eta_2(z=0)$  for relatively long  pulses of fundamental radiation.  (a) Input pulse area $S_{10}= 0.9900$; output pulse areas $S_{1L}=  0.2516$, $S_{20}= 0.3692$. (c)  Input pulse area $S_{10}= 0.9900$; output pulse areas $S_{1L}=0.0161$, $S_{20}= 0.4870$.}
\end{figure}
\begin{figure}[htbp]
\begin{center}
\includegraphics[width=.49\textwidth]{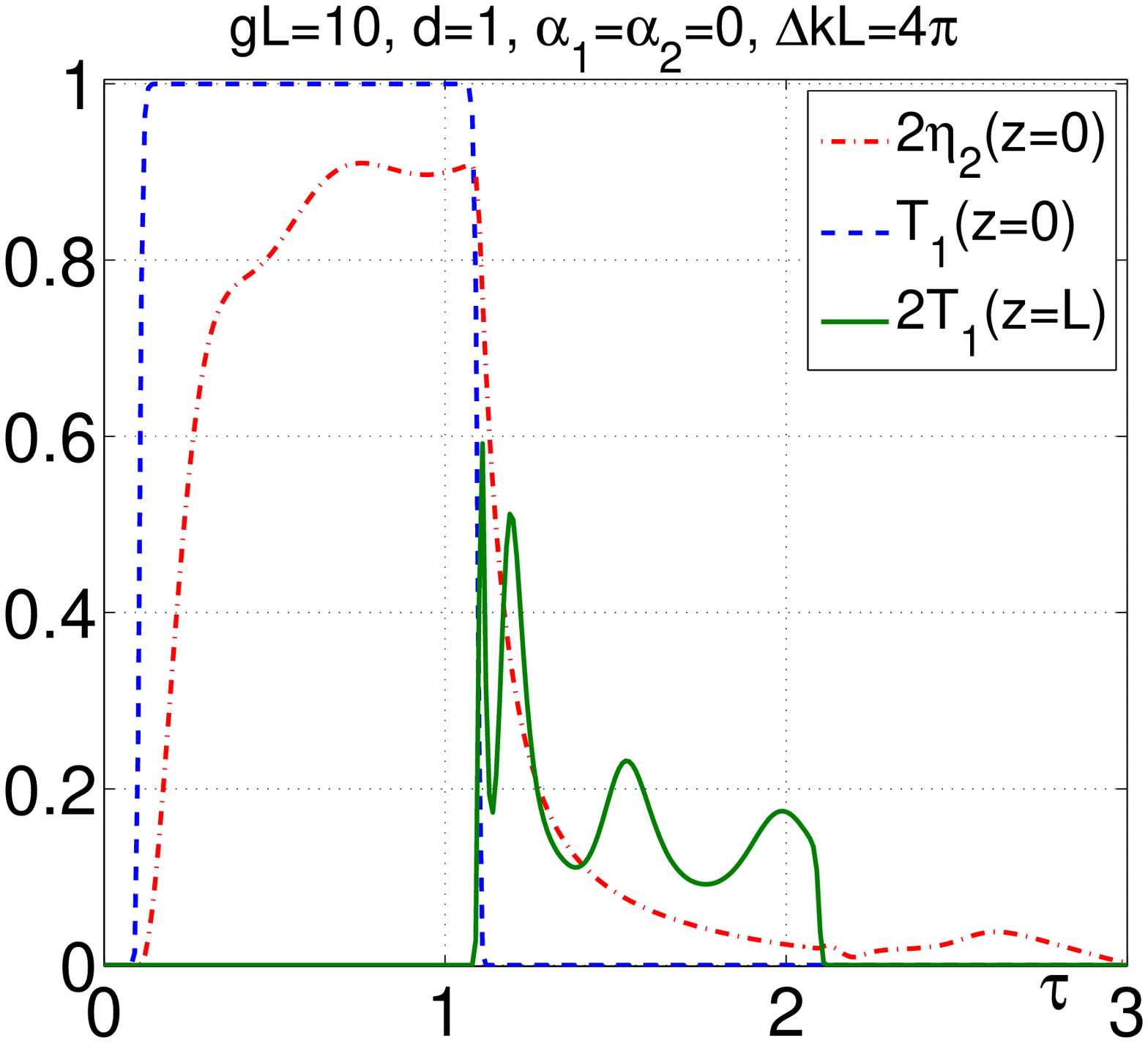}
\includegraphics[width=.49\textwidth]{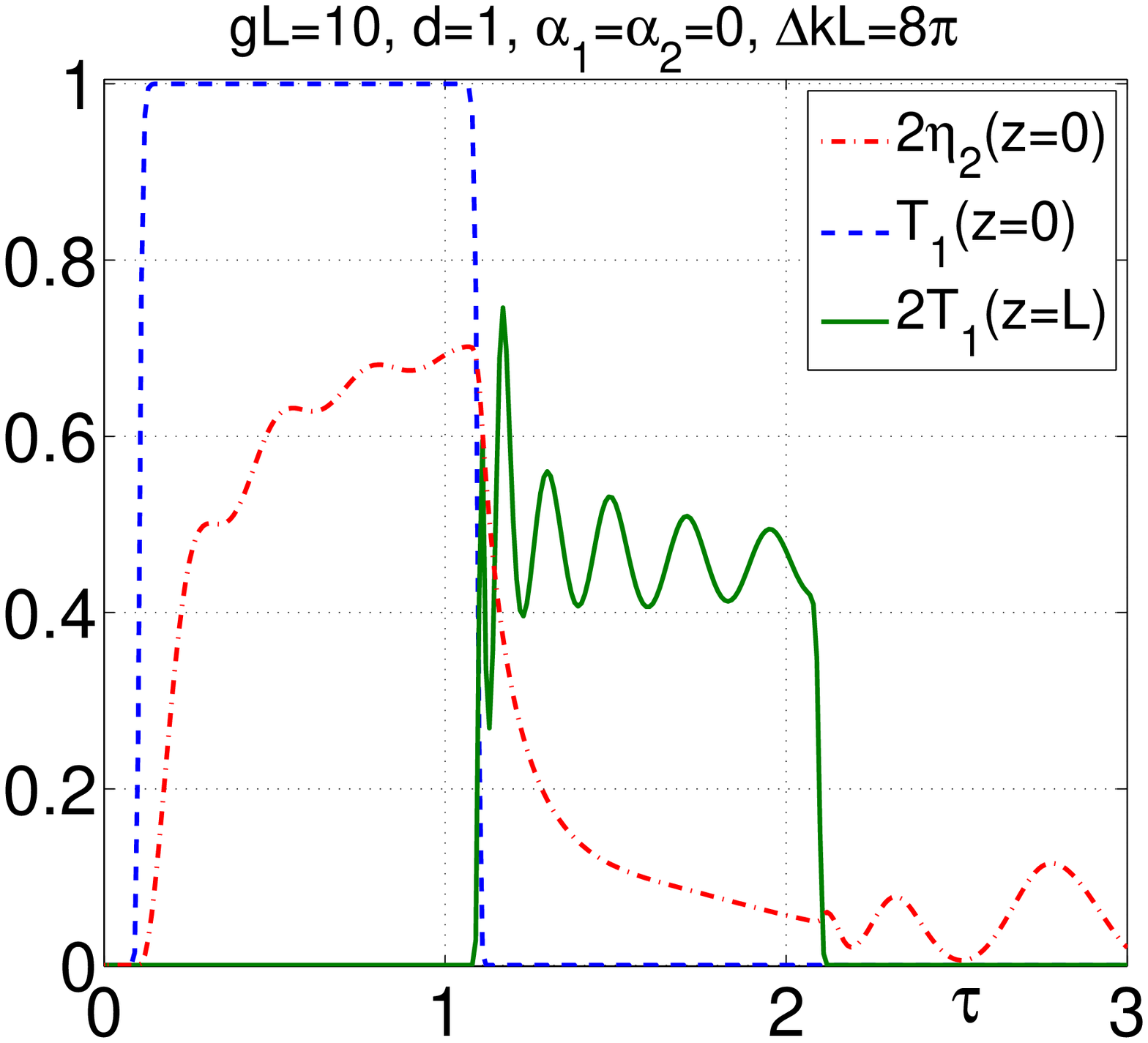}\\
(a)\hspace{80mm} (b)\\
\end{center}
\caption{\label{fi9} Effect of phase mismatch on SHG in a NIM slab for the case of strong, relatively long pulses.
}\end{figure}

\begin{figure}[htbp]
\begin{center}
\includegraphics[width=.6\textwidth]{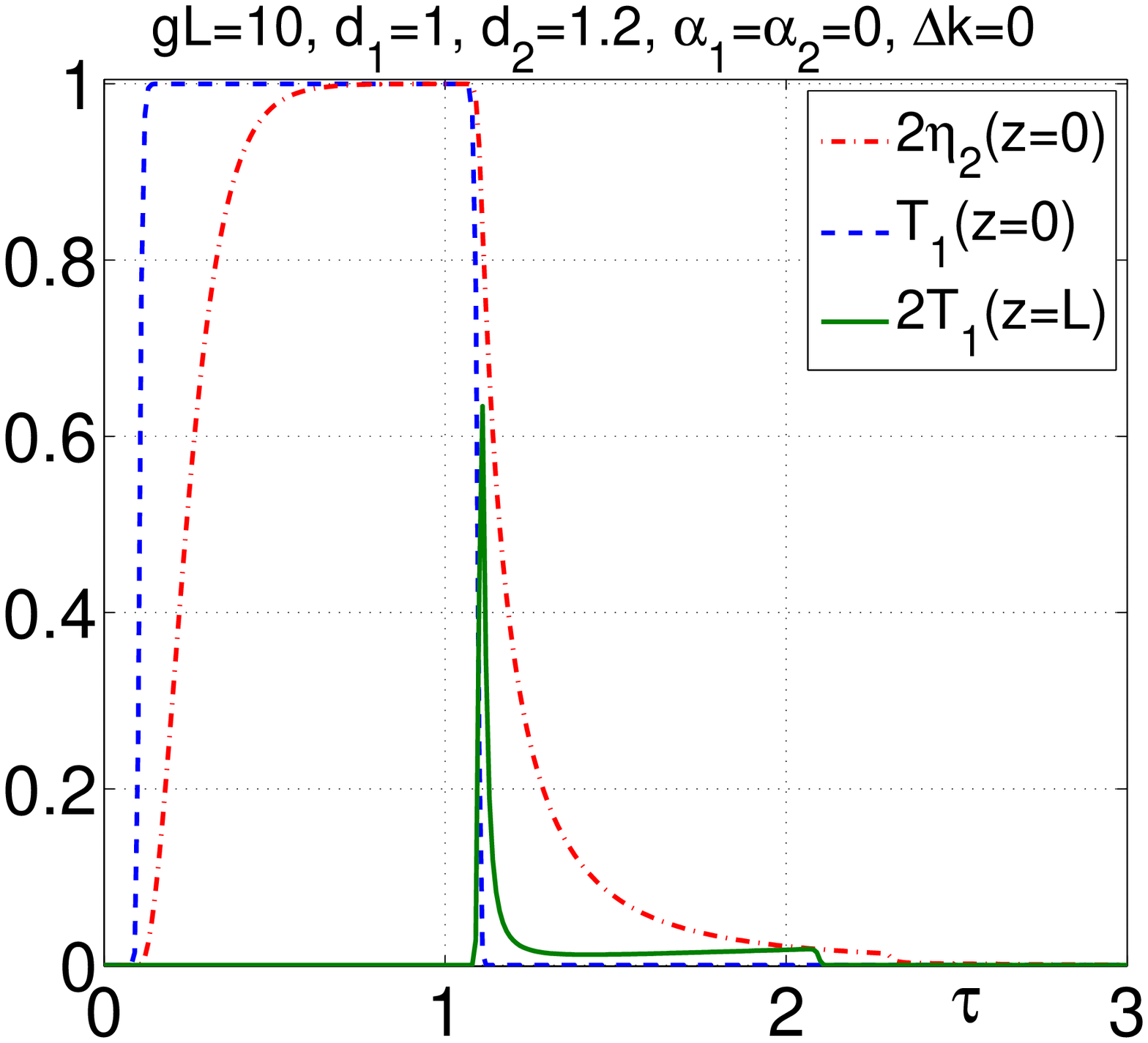}
\end{center}
\caption{\label{fi11}
Effect of difference in the group velocities on SHG in a NIM slab for the case of strong, relatively long pulses.
}
\end{figure}

The unusual properties of SHG in NIMs in the pulsed regime stem from the fact that it occurs only inside the pulse of fundamental radiation. Generation begins on the leading edge of the pulse, grows towards its trailing edge, and then exits the pulse with no further changes. Since the fundamental pulse propagates across the slab, the duration of the SH pulse is longer than the fundamental one. Depletion of the fundamental radiation along its pulse length and the conversion efficiency depend on its initial maximum intensity and phase matching. Ultimately, the overall properties of SHG, the pulse length, and the photon conversion efficiency depend on the ratio of the fundamental pulse and slab length. This extraordinary behavior is illustrated in Figs.~\ref{fi1}-\ref{fi11}.

Figures \ref{fi1}-\ref{fi11} display the input shape of the fundamental pulse $T_1=|a_1(z)|^2/|a_{10}|^2$ at $z=0$ when its leading front enters the medium and the results of numerical simulations  for the output fundamental pulse when its tail reaches the slab boundary at $z=L$ as well as  the shape and the conversion efficiency of the output  second harmonic pulse  $\eta_2=|a_2(z)|^2/|a_{10}|^2$ traveling against the z-axis, when its tail passes the slab's edge at $z=0$. For clarity, here, the medium is assumed loss-less and group velocities of the fundamental and SH pulses assumed equal for Figs.~\ref{fi1}-\ref{fi9}.

Figure \ref{fi1} corresponds to the fundamental pulse four time shorter than the slab thickness. It shows increase of the conversion efficiency  with increase the intensity of the input pulse. It is followed by the shortening of the SH pulse. Phase mismatch causes changes in the depletion and shape of the output fundamental pulse. However overall conversion rate does not change significantly. The outlined properties satisfy to the conservation law: the number of annihilated pair of photons of fundamental radiation ($S_{10}-S_{1L})/2$ is equal to the number of output SH photons $S_{20}$.

Figure~\ref{fi5} displays the corresponding changes for a longer input pulse length equal to 0.5 of the slab thickness. Here, conversion efficiency increases at a lower input intensity because of the longer conversion length. The changes in the SH pulse length and conversion efficiency with increasing input intensity appear less significant.

Figure~\ref{fi9} shows the effect of phase mismatch on the shape of the SH pulse for relatively long pulses and a high-intensity input pulse of fundamental radiation.

Figure~\ref{fi11} shows that the properties of the output SH pulse do not change significantly with an increase of the group velocity difference up to 20\% for long pulses and a high-intensity fundamental field.

\section{Three-wave mixing of contra-propagating electromagnetic waves: tailored transparency and reflectivity of a plasmonic metaslab, and applications to ultracompact nonlinear-optical data processing chips and sensors }\label{twm}
\begin{figure}[h!]
\begin{center}
\resizebox{.245\columnwidth}{!}{
\includegraphics{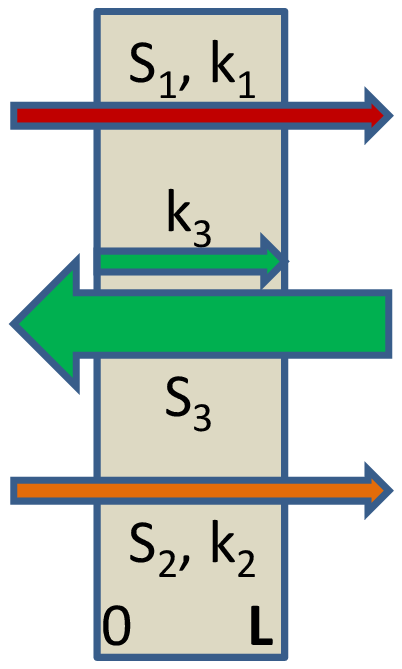}}
\resizebox{.24\columnwidth}{!}{
\includegraphics{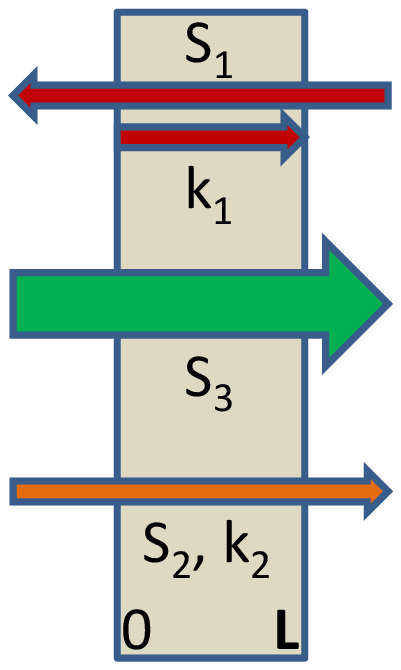}}
\resizebox{.245\columnwidth}{!}{
\includegraphics{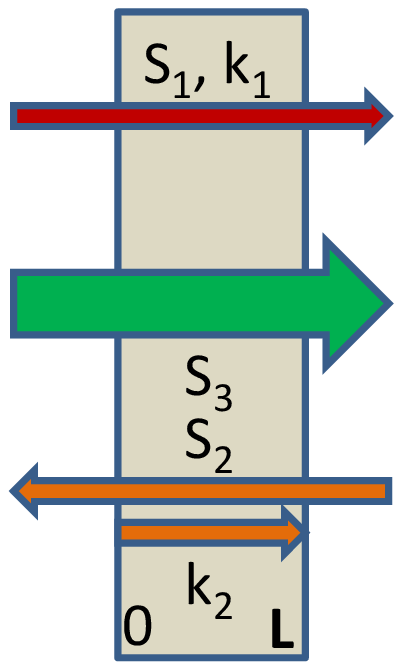}}\\
(a)\hspace{40mm}(b)\hspace{40mm}(c)\\
\caption{\label{fig4} Three different phase matching options for three-wave mixing of ordinary and backward light waves. (a)~$\mathbf{S}_{1,2}$ and $\mathbf{k}_{1,2}$  are energy fluxes and wavevectors  for the ordinary, positive index signal and generated idler; $\mathbf{S}_3$ and $\mathbf{k}_3$  -- for the negative index control field. (b), (c) Alternative schemes. (b) The NLO chip amplifies the negative-index signal traveling against the control beam [$n(\omega_1)<0$] and frequency upconverts it to the contra-propagating beam. (c) The NLO chip generates a negative-index idler and, therefore, shifts the frequency and reflects back a positive-index signal traveling along the control beam [$n(\omega_2)~<~0$].}
\end{center}
\end{figure}

An extraordinary electromagnetic property of NIMs stems from the fact that the energy flow and phase velocity of electromagnetic waves become counter-directed inside the NIM slab. Usually, a negative refractive index exists only inside a certain frequency band. The metamaterial remains ordinary, PI, outside that band. Consider a slab of thickness $L$ that possesses a quadratic nonlinearity, which enables three-wave mixing processes $\omega_3=\omega_1+\omega_2$. The amount of coherent energy transfer between the coupled light waves increases with the decrease of phase mismatch $\Delta kL$, where $\Delta k=k_3-k_2-k_1$. This dictates the requirement of co-directed wavevectors for all coupled waves. Consequently, the energy flux for the  wave that falls in the negative-index frequency domain appears contra-directed relative two others. Figure~\ref{fig4} depicts three corresponding possible options for the phase-matched TWM NLO coupling of ordinary and backward waves.

For the continuous-wave regime, the equations for the slowly varying amplitudes of the coupled waves can be written in the form
\begin{eqnarray}
da_1/dz&=&\pm[iX_1a^*_{2}a_3\exp(i\Delta k z)-({\alpha_1}/{2})a_{1}],\label{a1}\\
da_2/dz&=&iX_2a^*_{1}a_3\exp(i\Delta k z)-({\alpha_2}/{2})a_{2},\label{a2}\\
da_3/dz&=&\mp[iX_{3}a_1a_2\exp(-i\Delta k z)-({\alpha_3}/{2})a_{3}] \label{a3}.
\end{eqnarray}
The minus and plus signs refer to the negative- and positive-index waves, correspondingly. Here, $\Delta k=k_3-k_2-k_1$, and all three waves experience strong dissipation described by absorption indices $\alpha_{1,2,3}$. The slowly varying  effective  amplitudes of the waves, $a_{e,m,j}$, (j=\{1,2,3\}) and
nonlinear coupling parameters, $g_{e,m}$,  for the
{electric (\emph{e})} and magnetic (\emph{m}) types of quadratic nonlinearity  can be conveniently introduced as
\begin{eqnarray*}
a_{ej}=\sqrt{|\epsilon_j/k_j|}E_j,
X_{e}=\sqrt{|k_1k_2/\epsilon_1\epsilon_2|}  2\pi\chi^{(2)}_{ej};
\nonumber\\
a_{mj}=\sqrt{|\mu_j/k_j|}H_j,
X_{m}=\sqrt{|k_1k_2/\mu_1\mu_2|}  2\pi
\chi^{(2)}_{mj}.\nonumber
\end{eqnarray*}
The quantities  $|a_j|^2$ are proportional to the photon numbers in the energy fluxes. Equations for the amplitudes $a_j$ are identical for the both types of the nonlinearities.

We note the following three fundamental differences in Equations (\ref{a1}) and (\ref{a3}) as compared with their counterparts in ordinary, PI materials [Eq.~(\ref{a2})]. First, the signs with $g_{1}$ or $g_{3}$ are opposite to that with $g_{2}$. This is because the corresponding $\epsilon_{j}<0$ and $\mu_{j}<0$. Second, the opposite sign appears with $\alpha_{j}$ because the corresponding energy flow $\mathbf{S_{j}}$ is against the $z$-axis. Third, the boundary conditions for the negative-index wave are defined at the opposite side of the sample as compared to the ordinary waves because their energy flows are counter-directed. These modifications lead to dramatic changes in the equation solution as compared with the exponential dependence on $z$ that is characteristic for the ordinary,  naturally occurring, PI materials.

Consider the example depicted in Fig.~\ref{fig4}(a). We assume that the wave at $\omega_{1}$ with the wavevector $\mathbf{k}_1$ directed along the $z$-axis is a PI ($n_{1}>0$) signal. Usually it experiences strong absorption caused by metal inclusions. The medium is assumed to possess a quadratic nonlinearity $\chi^{(2)}$ and is illuminated by the strong, higher-frequency control field at $\omega_{3}$, which falls into the NI domain. Due to the TWM interaction, the control and signal fields generate a difference-frequency idler at $\omega_{2}=\omega_{3}-\omega_{1}$, which is also assumed to be a PI wave ($n_{2}>0$). The idler, in cooperation with the control field, contributes back into the wave at $\omega_{1}$ through the same type of TWM interaction and thus enables OPA at $\omega_{1}$ by converting the energy of the control fields into the signal. In order to ensure effective energy conversion the induced traveling wave of nonlinear polarization in the medium and the coupled electromagnetic wave at the same frequency must be phase matched. Hence, all phase velocities (wavevectors) must be co-directed. Since $n(\omega_3)<0$, the control field is a backward wave, i.e., its energy flow $\mathbf{S}_{3}
=(c/4\pi)[\mathbf{E_3}\times \mathbf{H_3}]$ is directed against the $z$-axis. Such a device can be employed as NLO sensor. It can be conveniently and remotely interrogated to actuate frequency up-conversion and amplification of a signal directed towards the remote detector. Such a signal could be incoming far-infrared thermal radiation emitted by the object of interest or signal that carries important spectral information about the chemical composition of the environment, for example. Two other schemes depicted in Fig.~\ref{fig4}(b),(c) offer different advantages and operational properties for NLO NIM-based devices.
The research challenge is that such an unprecedented NLO coupling scheme leads to changes in the set of coupled nonlinear propagation equations and boundary conditions compared to the standard ones known from published literature. This in turn results in dramatic changes in their solutions and in the multi-parameter dependencies of the operational properties of the proposed devices.

Unusual propagation and energy-conversion properties in double-domain NIM/PIM slabs are readily seen, for example, in the coupling scheme of Fig.~\ref{fig4}(a) and with a loss-free medium. Then the following Manley-Rowe relations can be derived from Maxwell's equations for the slowly varying amplitudes:
\begin{eqnarray}\label{manley}
&{d}(|a_1|^2-|a_2|^2)/{dz}=0,& \label{a12}\\
&{d}(|a_3|^2-|a_1|^2)/{dz}=0, \quad
{d}(|a_3|^2-|a_2|^2)/{dz}=0.&\label{a23}
\end{eqnarray}
Here, $|a_j|^2$ present photon numbers. Equation~(\ref{a12}) predicts  that the difference of the numbers of  photons $\hbar\omega_1$ and  $\hbar\omega_1$ remains constant through the sample, which indicates their creation in pairs due to the split of photons $\hbar\omega_3$. However, Eqs.~(\ref{a23}) predict  that the differences of the numbers of  photons $\hbar\omega_1$ and  $\hbar\omega_3$ as well as of $\hbar\omega_2$ and  $\hbar\omega_3$ \emph{also remain constant} through the sample. This looks like a breaking of the energy conservation law and is in seemingly striking difference with the fact that the \emph{sum} of the corresponding photon numbers is constant in the analogous case in a PIM. Actually, such unusual dependencies stem from the fact that the waves propagate in the opposite direction. Consequently, \emph{extraordinary distributions} of these fields across the slab and the dependence of their output values on the linear and nonlinear optical properties of the given NIM and on the input intensities of the coupled fields are expected, especially when the conversion efficiency becomes large. Particularly, the conversion rate is expected to grow across the slab with a different rate than in an ordinary medium with a standard coupling geometry. Equations~(\ref{a23}) indicate unusual feedback, which provides correlated depletion of the control field on one hand and growth of the signal and the idler on the other hand, so that the difference must remain constant along the metaslab. Absorption would change this behavior, which may strongly depend on the absorption dispersion and on the phase mismatch. Investigation of the indicated dependencies is important for optimizing the operational properties of the proposed sensor. Consequently, accounting for the boundary conditions that must be defined at the opposite facets of the slab, this suggests extraordinary distributions of these fields across the slab and extraordinary dependencies of their output values on the linear and nonlinear optical properties of the given NIM and on the input magnitudes of the fields. Such dependencies are essentially different for the coupling schemes depicted in Fig.~\ref{fig4}.

\begin{figure}[h!]
\begin{center}
\includegraphics[width=.45\columnwidth]{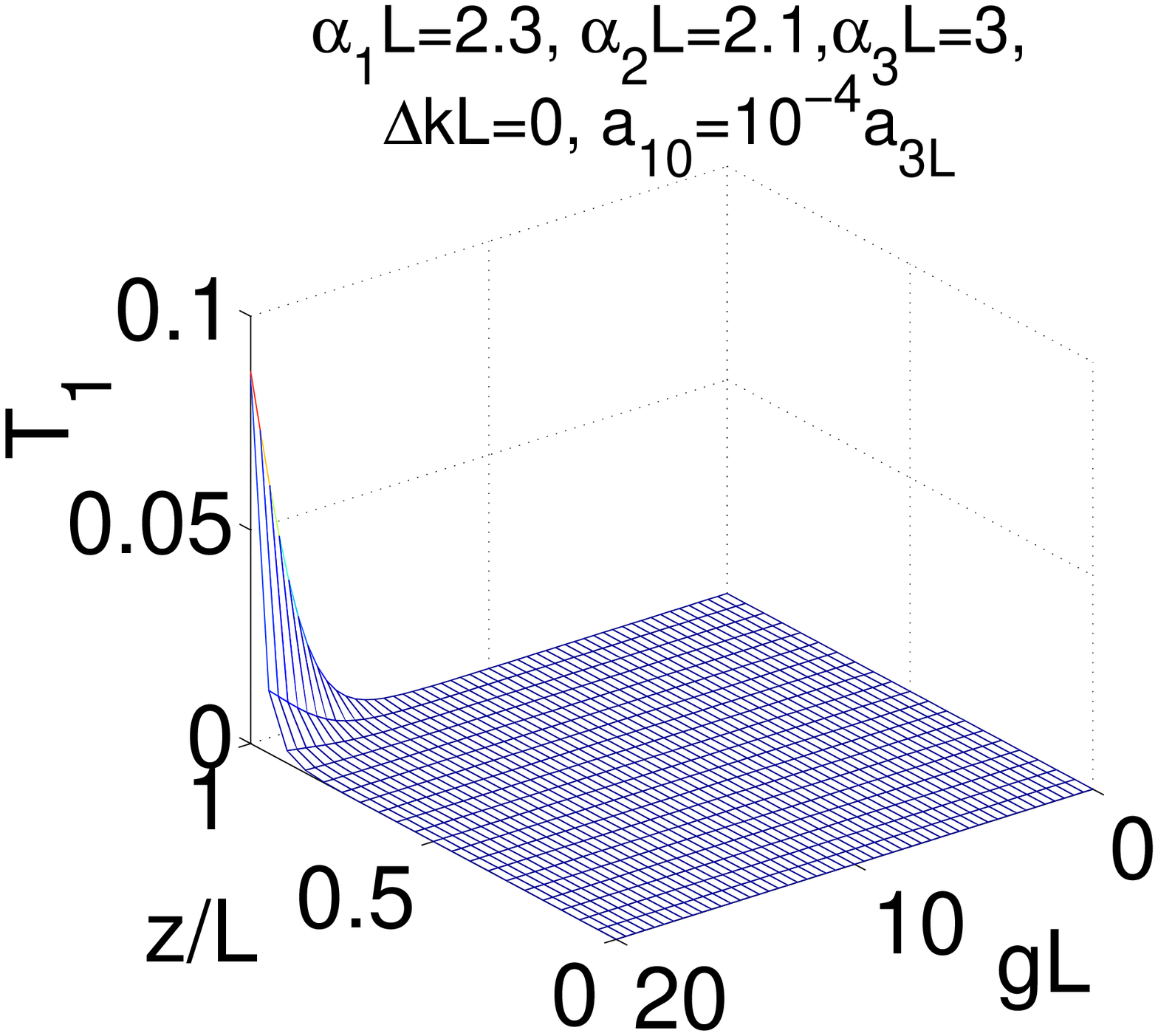}
\includegraphics[width=.45\columnwidth]{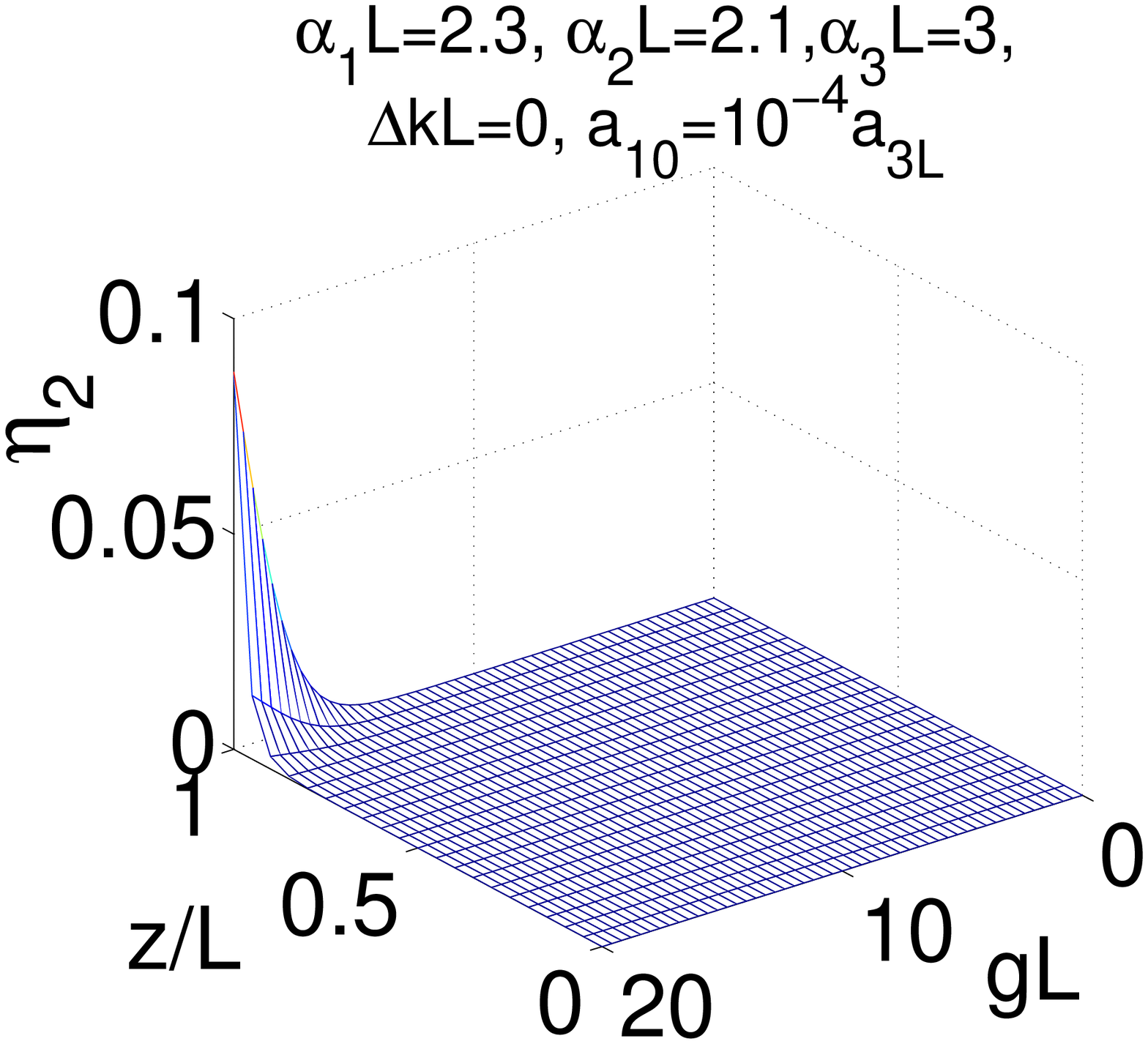}\\
\includegraphics[width=.45\columnwidth]{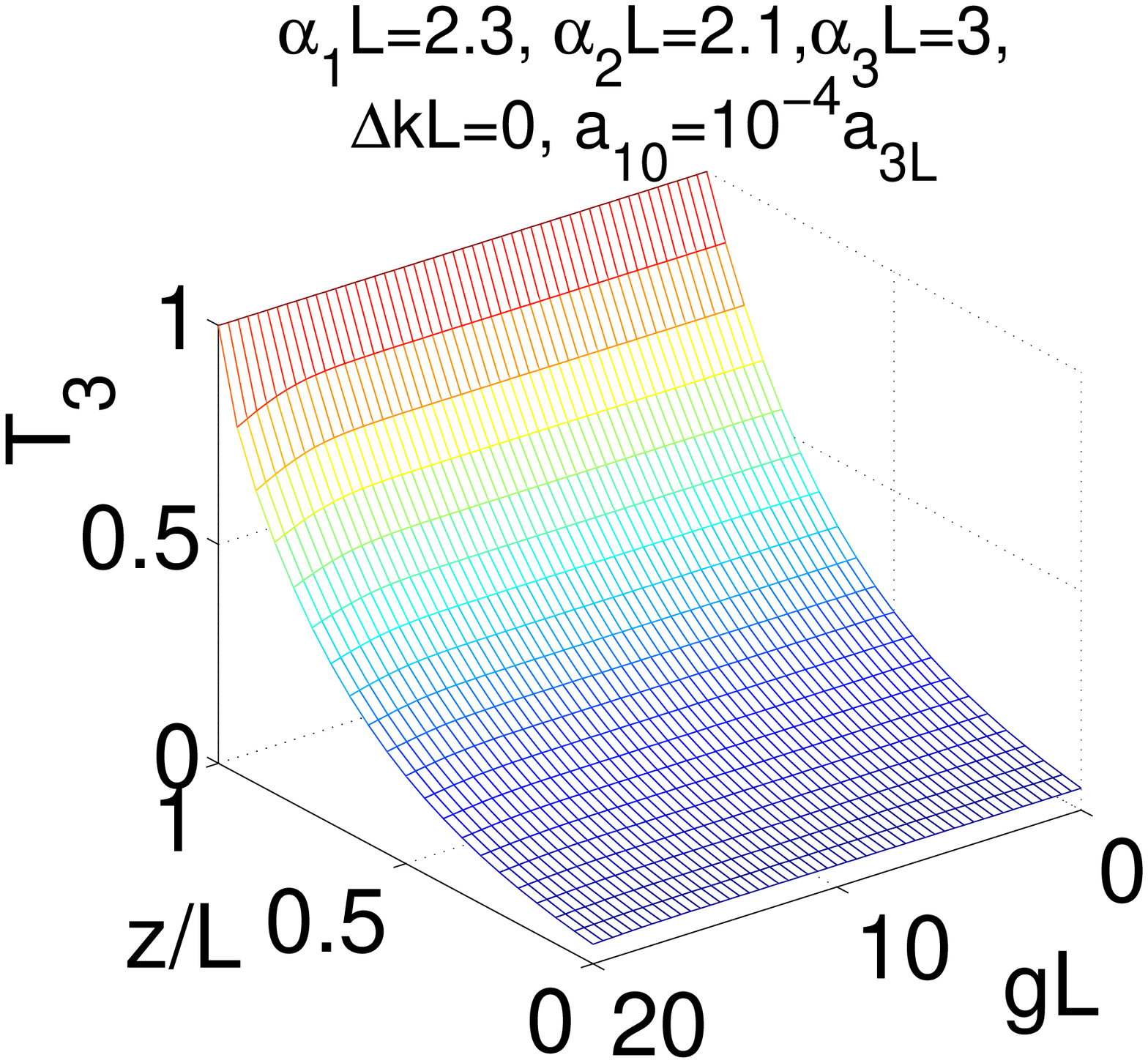}
\caption{\label{opa_weak_w_abs} Distribution of the coupled fields across the slab and output characteristics of the amplified transmitted, generated, and depleted control fields for the case of a weak input signal $a_{10}=10^{-4}a_{3L}$. }
\end{center}
\end{figure}
\begin{figure}[h!]
\begin{center}
\includegraphics[width=.45\columnwidth]{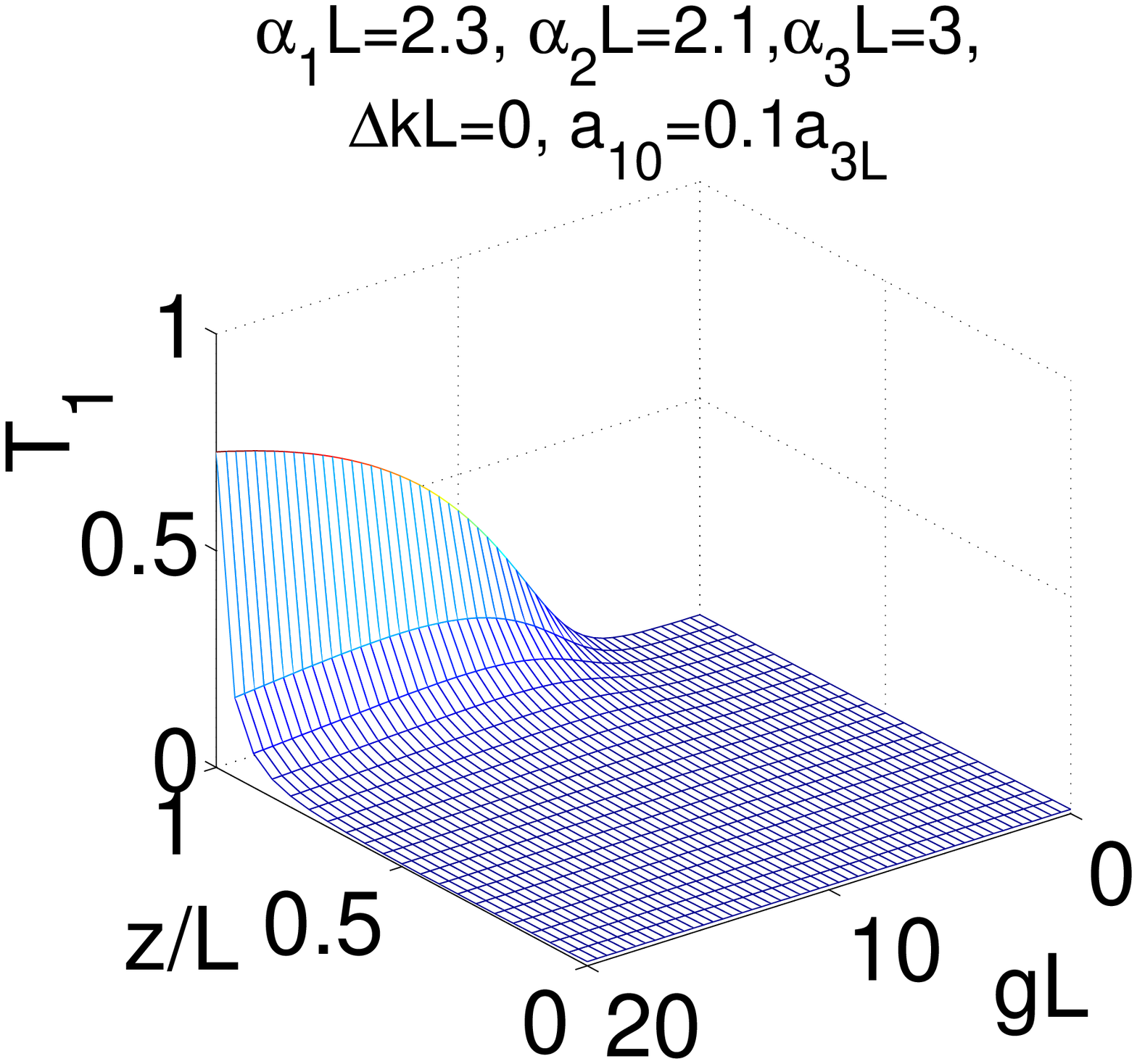}
\includegraphics[width=.45\columnwidth]{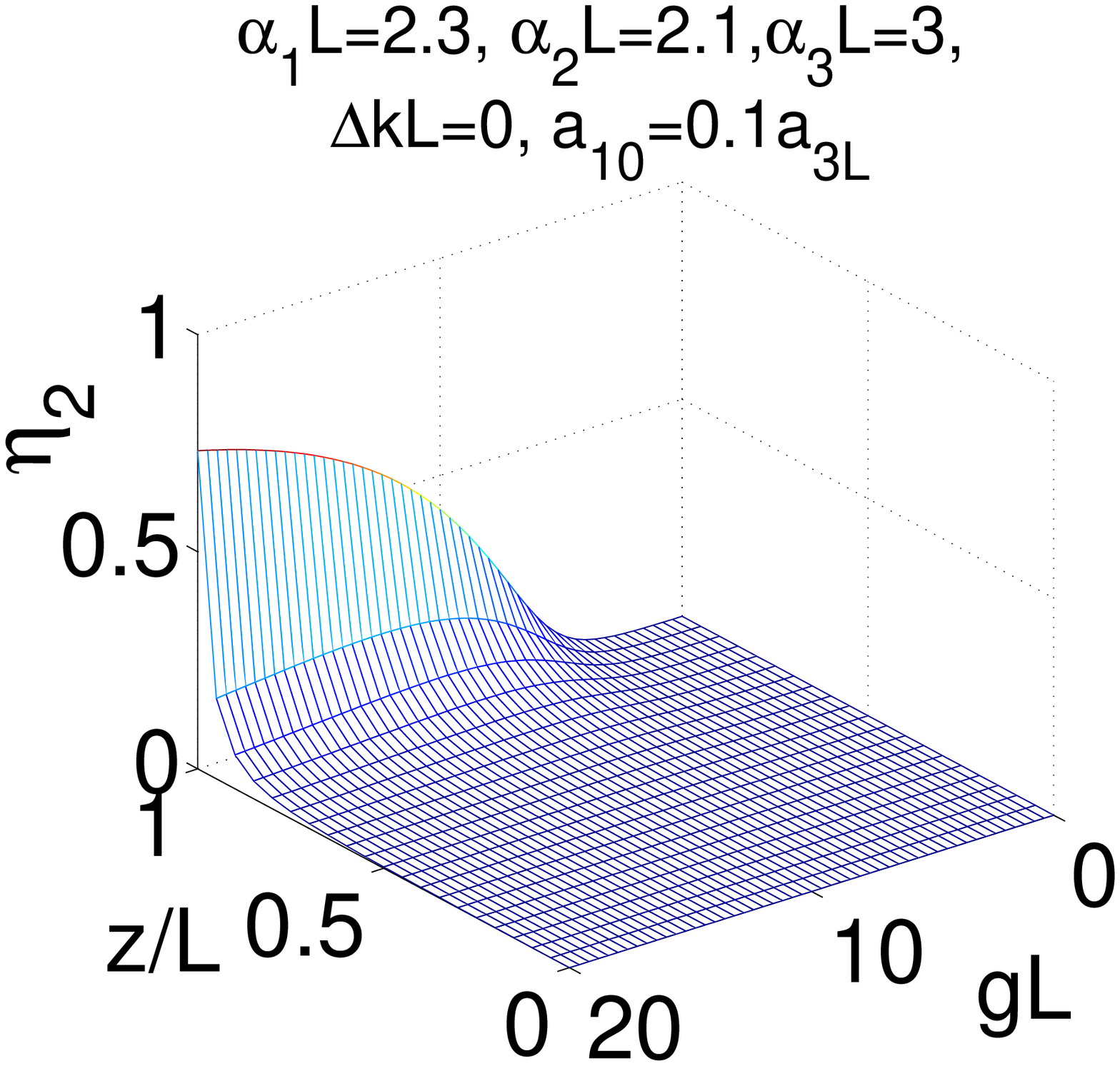}\\
\includegraphics[width=.45\columnwidth]{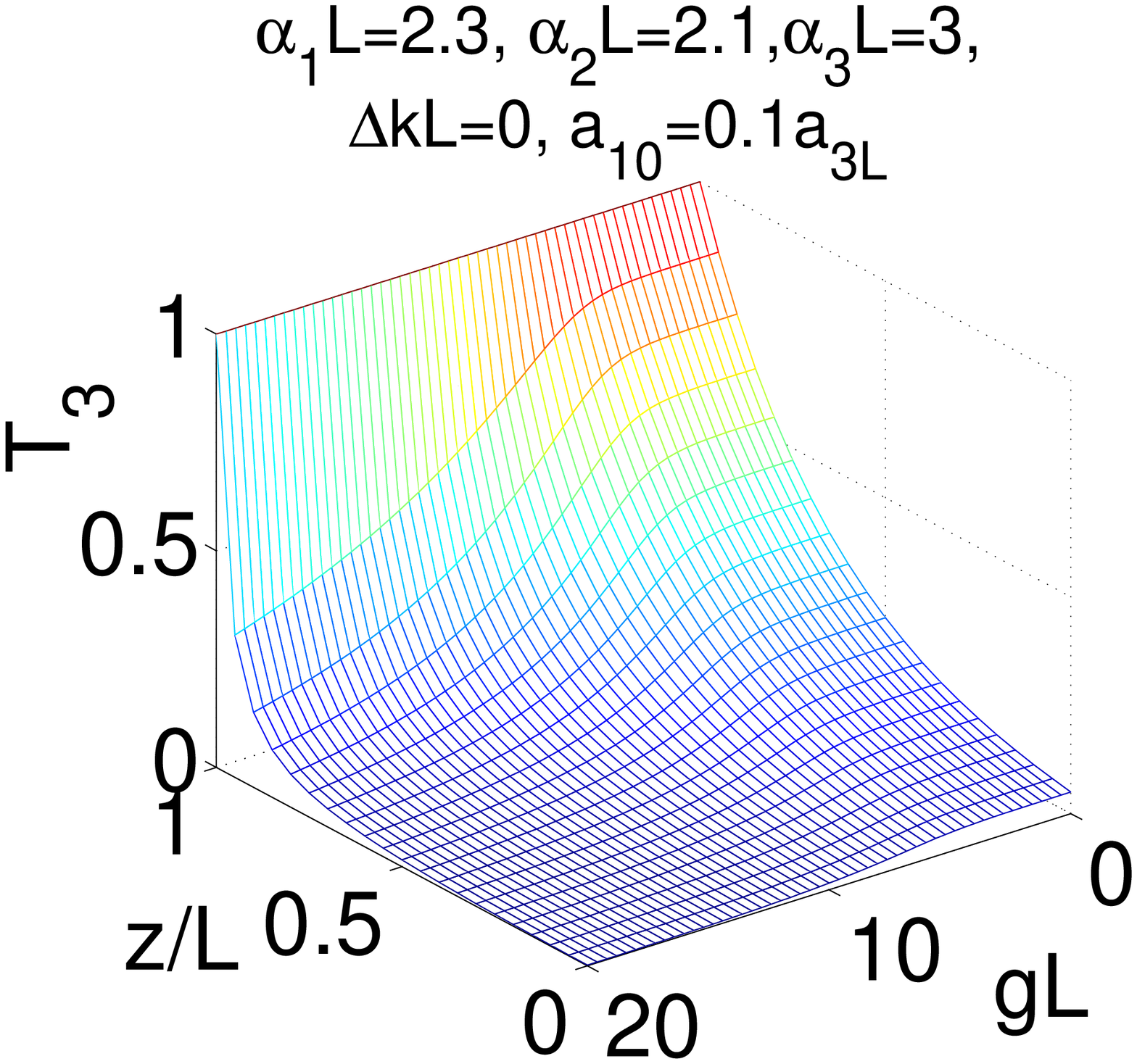}
\caption{\label{opa_strong_w_abs} Distribution of the coupled fields across the slab and output characteristics of the amplified transmitted, generated, and depleted control fields for the case of a strong input signal $a_{10}=0.1 a_{3L}$.}
\end{center}
\end{figure}
\subsection{Properties of a nonlinear optical reflector and amplifier tailored by the negative-index control field}\label{mcon}
Consider the scheme depicted in Fig.~\ref{fig4}(a). An analytical solution is not possible for the problem that underlies the concept of the proposed sensor.
Figures \ref{opa_weak_w_abs} and \ref{opa_strong_w_abs} display the results of numerical simulations for this case. Here, z is the length across the slab of thickness L, $T_3(z)=|a_3(z)/a_{3L}|^2$ and $T_1(z)=|a_1(z)/a_{3L}|^2$ are transparencies, $\eta_2(z)=|a_2(z)/a_{3L}|^2$ is the photon conversion efficiency, $g=Xa_3(L)$, $X_{ej}=\sqrt{|k_1k_2/\epsilon_1\epsilon_2|}  2\pi\chi^{(2)}_{ej}$, and $a_{3L}=a_3(L)$ is the amplitude of the input control field.  Absorption indices, $\alpha_j$, for the coupled field are indicated, and $\Delta k=0$. Figure~\ref{opa_weak_w_abs} illustrates the case of a weak input signal so that the depletion of the control field due to the conversion becomes significant only in the vicinity of z=L. Figures~\ref{opa_weak_w_abs}(a) and (b) show the possibility to achieve many orders of amplification of the signal traveling against the control beam and its conversion to the frequency-shifted wave for the intensity of the incoming control field corresponding to $gL\approx 15...20$. Then the numbers of  the output photons $\hbar\omega_1$ and $\hbar\omega_2$ make about 10\% of that of the input contra-directed control field, which means amplification on the order of $10^7$.
Figure~\ref{opa_strong_w_abs} demonstrates a stronger energy-conversion effect due to a higher input intensity of the signal which, however, leads to lower overall amplification.

\subsection{Tunable NLO mirror employing a negative-index idler: weak signal and idler approximation }\label{mid}
\begin{figure}[h!]
\begin{center}
\resizebox{0.35\columnwidth}{!}{
\includegraphics{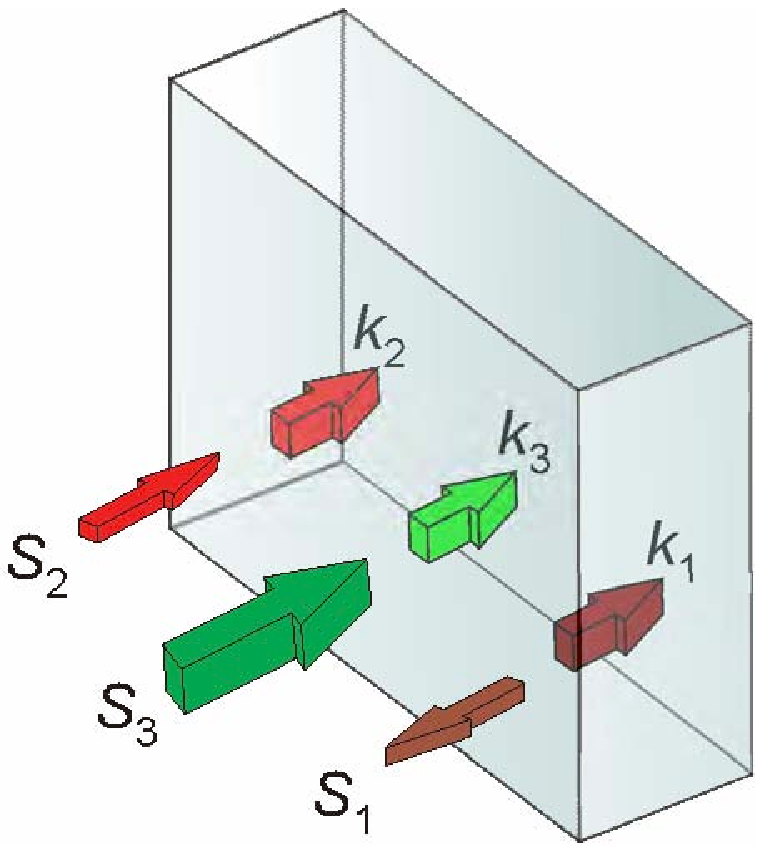}}
\resizebox{0.45\columnwidth}{!}{
\includegraphics{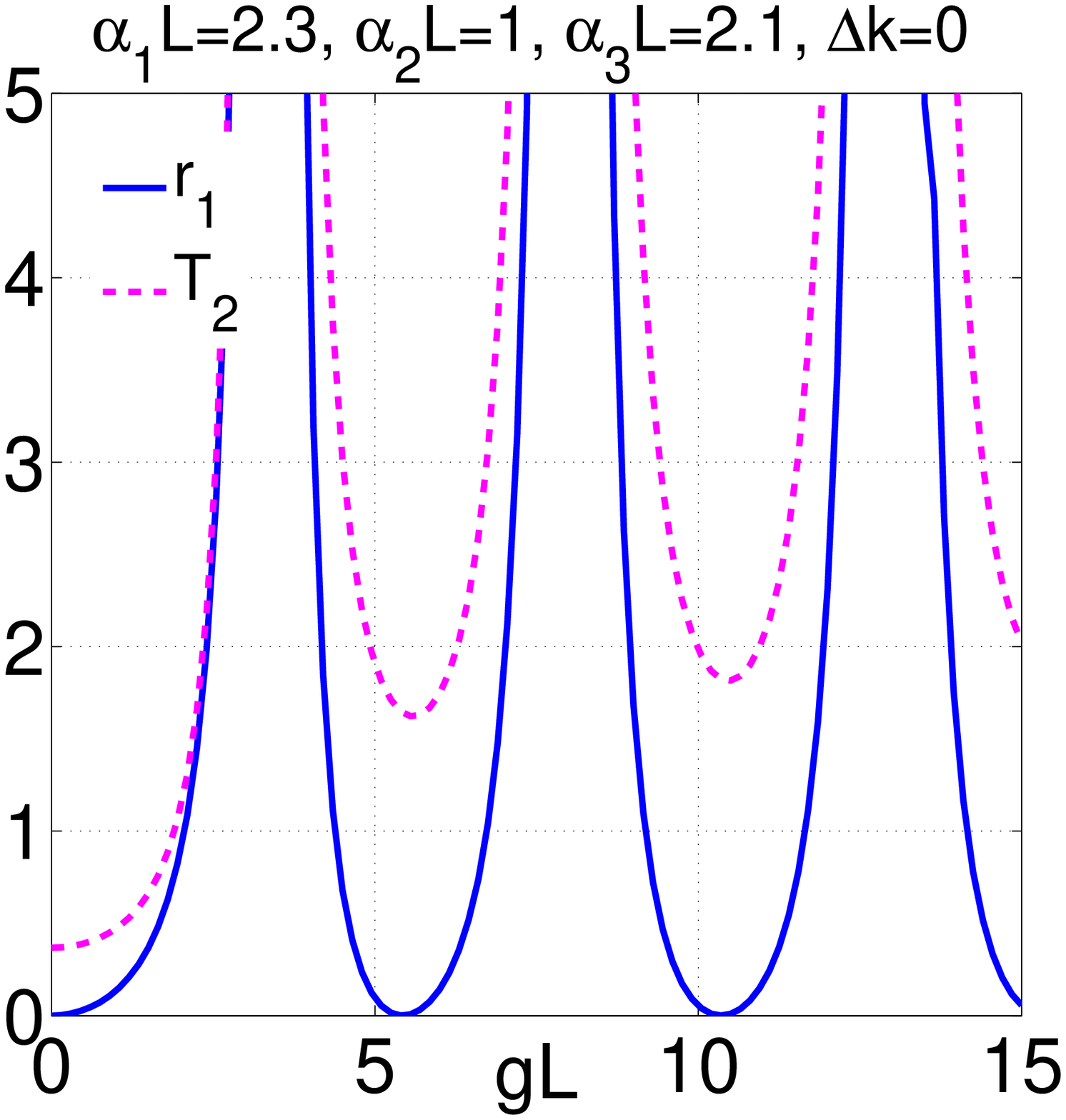}}\\
(a)\hspace{80mm}(b)\\
\end{center}
\caption{\small (a) Coupling scheme: $\mathbf{S}_2$ -- incident beam, $\mathbf{S}_3$ -- control beam, $\mathbf{S}_1$ -- generated negative-index (reflected) beam. (b) Characteristic dependence of reflectivity, $r_1$,  and transmittance, $T_2$,  of the NLO mirror on  the product of nonlinear susceptibility, strength of the control field and thickness of the NIM slab  with account for absorption of all coupled fields. Here, reflectivity can be switched from zero to  magnitudes exceeding 100\%.}
\label{fig5}
\end{figure}

The physical principles of the proposed nonlinear-optical micromirror, which also can be viewed as an optical data-processing chip, are based on difference-frequency generation of a backward, NI
wave in a strongly absorbing double-domain NIM slab [Fig.~\ref{fig5}(a), which is equivalent to Fig.~\ref{fig4}(c)].
Simultaneously, the incident PI signal propagating through the slab experiences OPA. Only the reflected-wave frequency is assumed to fall in the negative-index frequency domain. Here, two incident co-propagating PI waves would produce an all-optically controlled reflectivity accompanied by a tunable frequency shift of the reflected beam. A strong control field at $\omega_3$ and a strong incident wave at $\omega_2$ are both assumed to be PI. The generated difference-frequency wave at $\omega_1=\omega_3-\omega_2$ is NI and, therefore, a backward wave.  All three waves experience strong dissipation described by absorption indices $\alpha_{1,2,3}$.

First, consider a simplified model that neglects depletion in the control field due to the TWM conversion and absorption. Hence, the field is assumed to be uniform across the slab. This model allows one to understand the basic properties of TWM coupling of backward waves and difference-frequency generation of a backward NI wave. For the case of a spatially homogeneous, constant control field and a real, nonlinear susceptibility, an analytical solution to  Eqs.~(\ref{a1})-(\ref{a2}) can be found, and then the reflectivity, $r_{1}=|a_1(0)/a_{20}^{\ast}|^2$, is given by the equation
\begin{equation}
r_{1}=\left|\frac{(g/R)\sin RL}{\cos RL+\left(s/R\right) \sin RL}\right|^2,\quad  \label{gen11}
\end{equation}
whereas the corresponding transmission factor for the PI incident wave, $T_{2}=\left\vert
{a_{2}(L)}/{a_{20}}\right\vert ^{2}$, is
\begin{equation}
T_{2}=\left|\frac{\exp \left\{-\left[ \left(
\alpha_{2}/2\right)-s%
\right] L\right\}}{\cos RL+\left(s/R\right) \sin RL}\right|^2.
\label{T2}
\end{equation}
Here, $L$ is the thickness of the slab,
$g=g_0=Xa_{30}$, $R=\sqrt{g^{2}-s^{2}}$, and $s=[({\alpha_{1}+\alpha_{2}})/{4}]-[i{\Delta
k}/{2}]$. It is seen that the reflectivity and transmittance experience a \emph{set of ``geometrical'' resonances} \cite{Yar}. For example, at $s=0$, $r_{1}$=tan$^2(gL)$,  $r_{1}=1/cos^2(gL)$ and they tend to infinity at $g L \rightarrow (2j+1)\pi/2$, (j=0, 1, 2, ...), which indicates the possibility of \emph{mirrorless self-oscillations}. A similar behavior is characteristic for distributed-feedback lasers and is equivalent to a great extension of the NLO coupling length. It is known that even weak amplification per unit length may lead to lasing provided that the corresponding frequency coincides with high-quality cavity or feedback resonances. Such a \emph{giant enhancement} of the conversion efficiency in the vicinity of ``geometrical'' resonances is in \emph{striking contrast} with the exponential dependencies $ T_2\propto \exp(2gL)$ known for the counterparts in ordinary nonlinear materials.
Figure~\ref{fig5}(b) displays such resonances in an initially strongly opaque slab. It shows the dependence of the reflectivity $r_1$ and transmittance $T_2$ of the slab on the parameter $gL$ for specific absorption indices at the frequencies of the coupled waves indicated in the figure and $\Delta k=k_{3}-k_{2}-k_{1}=0$. It is seen that the modulation properties of the reflected and transmitted beams differ greatly. The transmitted beam experiences amplification for any magnitude of the parameter $gL$ above a certain level. On the contrary, the intensity of the reflected beam at $\omega_1$ can be varied over a wide range by changing the intensity of the control field. It can be modulated from zero to magnified values \emph{exceeding generation threshold} for both coupled waves.  The Manley-Rowe equations for the given coupling scheme predict \emph{the sum} of $|a_1|^2$ and $|a_2|^2$ to be conserved across the slab. Note that in a transparent PI medium, the \emph{difference} of these values is invariant across the slab. It appears that absorption of the control field causes broadening the resonances but does not destroy the resonance behavior. The transmission minima are dependent on the ratio of absorption rates; this is in contrast with the reflectivity minima, which remain robust. Ultimately, the NLO coupling under consideration may provide for significantly sharper growth of the entangled counter-propagating photons in the vicinity of the ``geometrical'' resonances than the conventional schemes for the same parameters of the slab and the control field. The described extraordinary properties can be utilized for sensing, filtering and conversion of weak light signals.

\begin{figure}[h]
\begin{center}
\resizebox{0.45\columnwidth}{!}{
\includegraphics{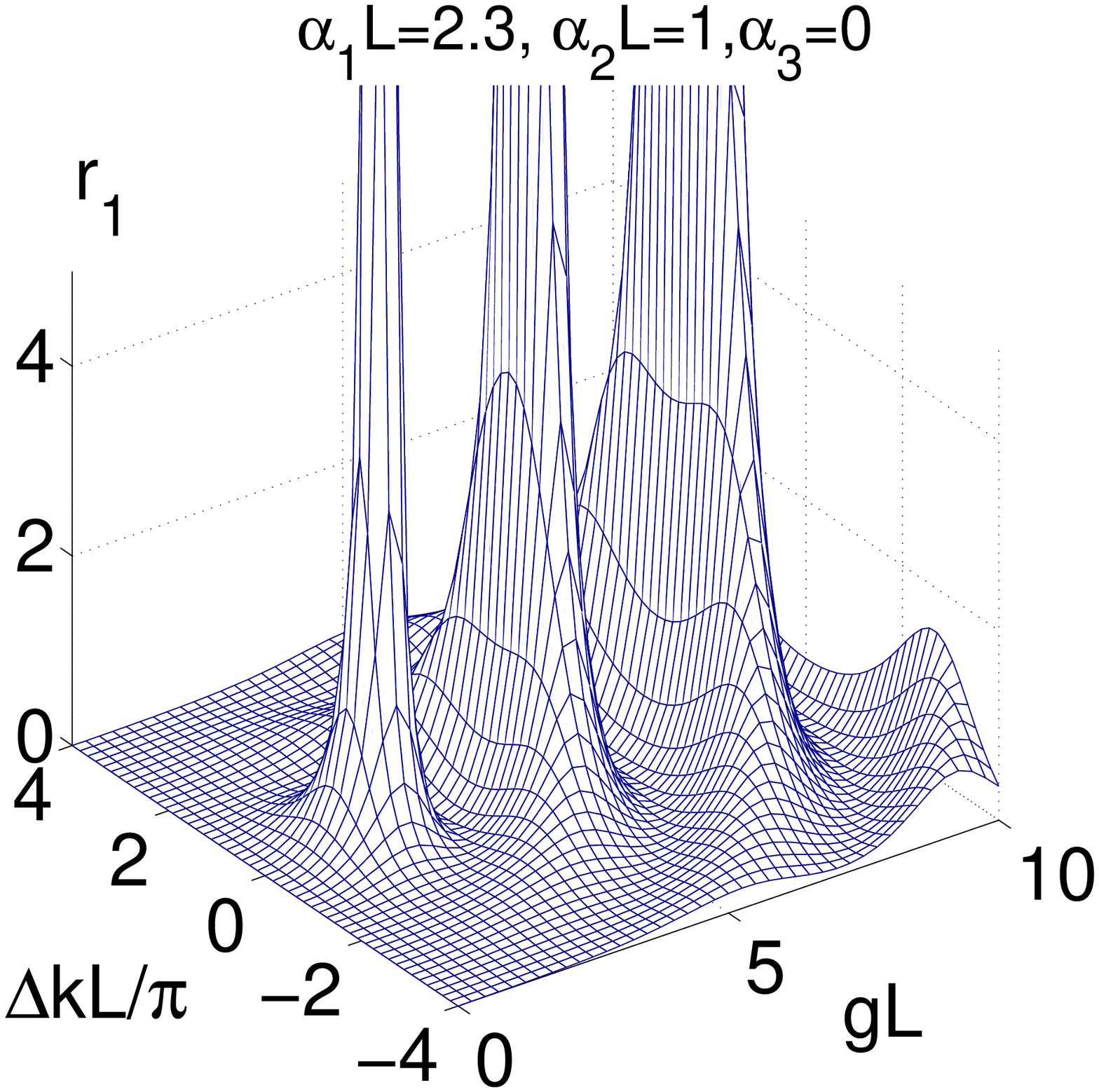}}
\resizebox{0.45\columnwidth}{!}{
\includegraphics{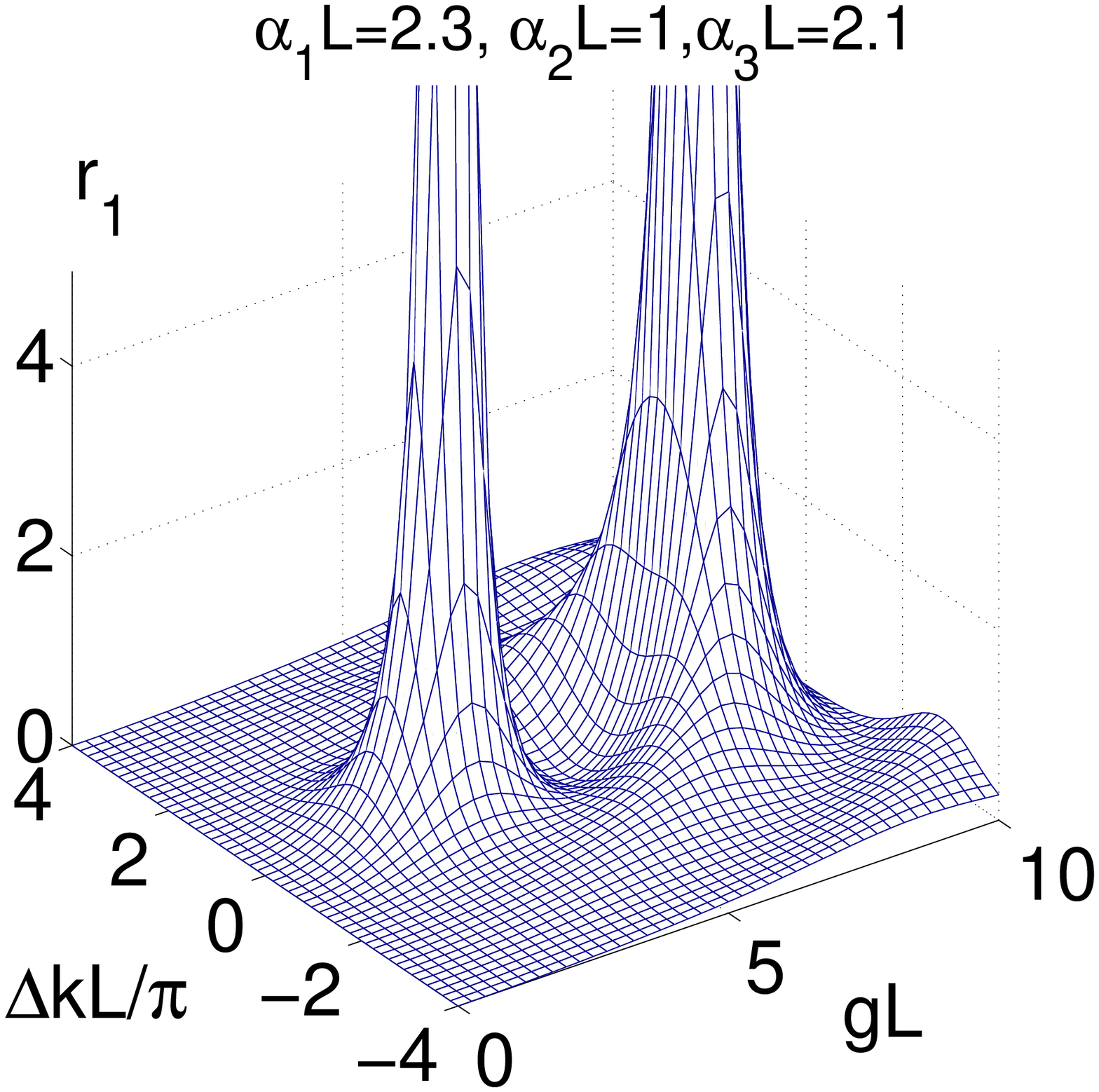}}
(a)\hspace{35mm}(b)\\
\resizebox{0.45\columnwidth}{!}{
\includegraphics{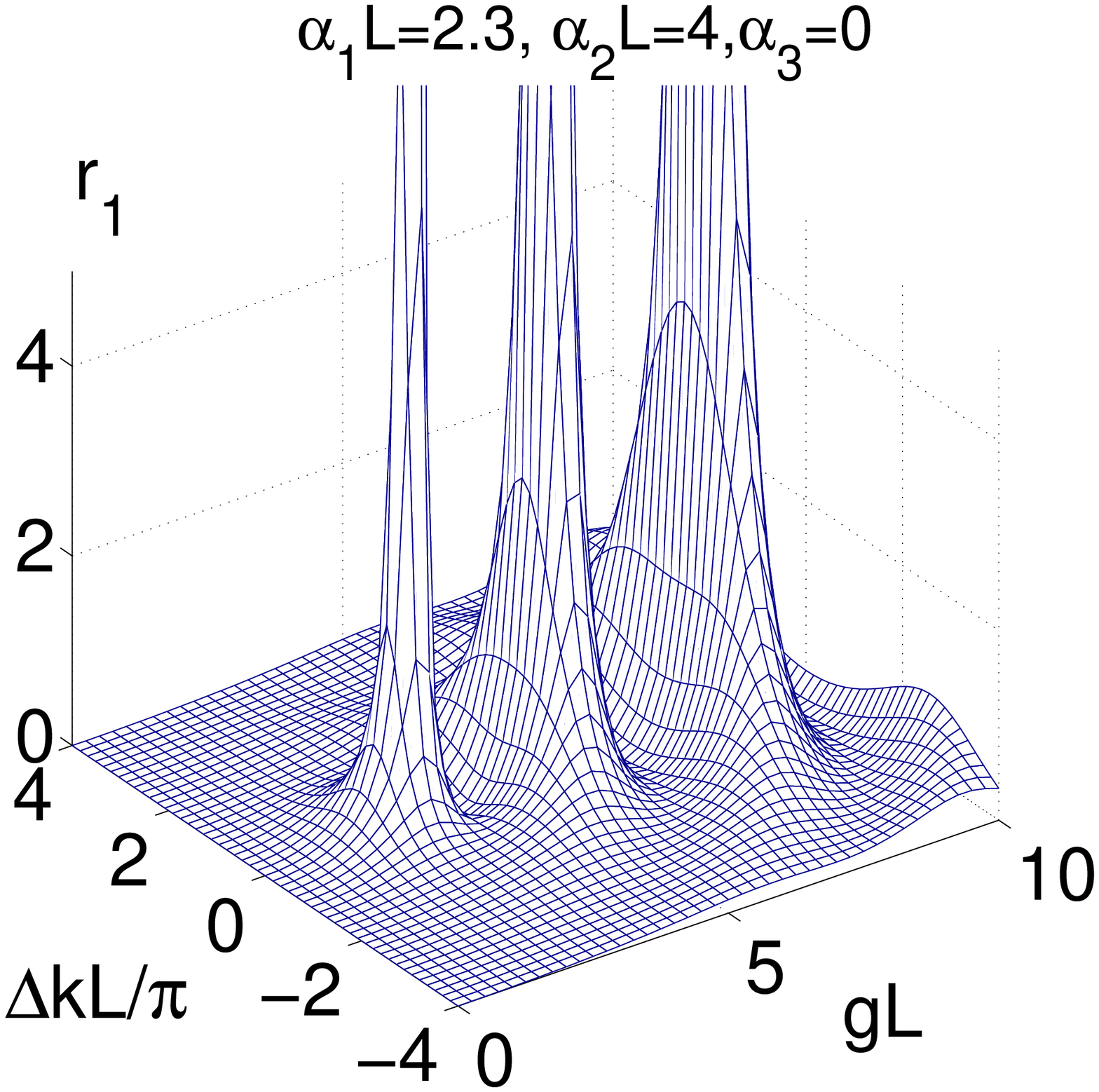}}
\resizebox{0.45\columnwidth}{!}{
\includegraphics{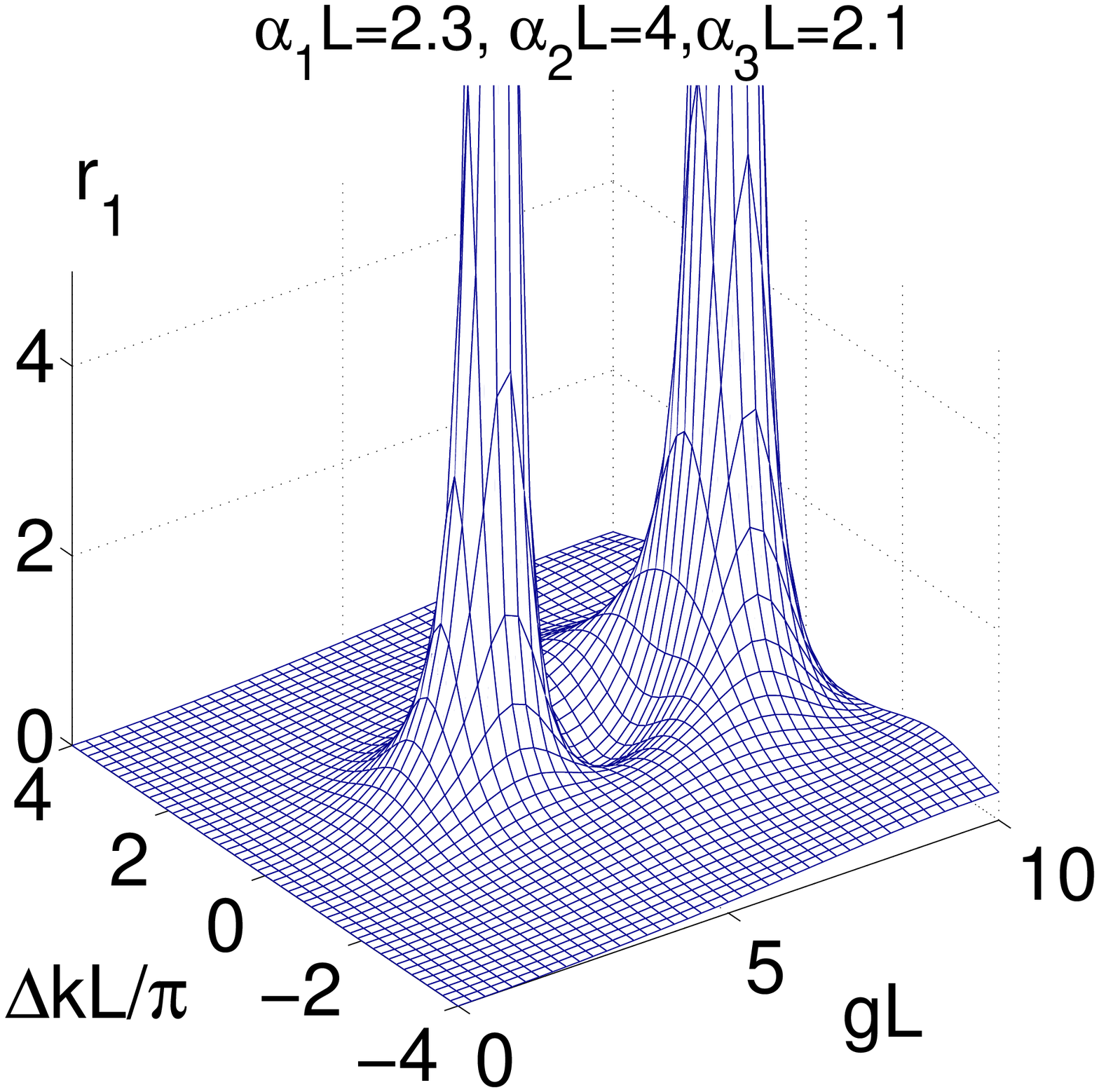}}
(c)\hspace{35mm}(d)
\caption{\label{f10} Reflectivity vs. intensity of the control field and phase mismatch for different absorption indices for the coupled waves.}
\end{center}
\end{figure}
Alternatively, the phase mismatch causes a decrease of the reflectivity maxima and an increase of the minima.  Reflectivity becomes relatively robust against phase mismatch with an increase of the intensity of the control field. It drops dramatically in the range of small phase mismatch and then remains relatively robust at the lower level within the range of greater phase mismatch, as seen in Fig.~\ref{f10}. The outlined properties of the NLO mirror are determined by the interplay of several processes that have a strong effect on the NLO coupling of contra-propagating waves and, consequently, on their distributions inside the slab.

The depletion of the control field due to energy conversion to the reflected and transmitted beams may have a strong effect on the processes under investigation. Particularly, it limits the amplification in the maxima and may even change the resonance behavior. Basically,  the reflected wave has a different frequency than the incident beam. The simulations predict that the quantum conversion efficiency with respect to the control field may be up to several tens of percent, which indicates great enhancement of the reflected and transmitted beams with respect to the incident, positive-index, weak field and the reflectivity that may significantly exceed 100\%. Ultimately, the simulations show  the possibility of \emph{all-optical tailoring and switching of the reflectivity and transparency} of such a mirror over a wide range by changing the intensity of the control field.
More details regarding the given coupling scheme and the one displayed in Fig.~\ref{fig4}(b) can be found in Refs. \cite{OL,APB,EPJD,WAS,Ch}

\subsection{Estimates}\label{est1}
The characteristic magnitude of the parameter $gL$, which is required to realize the effects predicted above, is on the order of 1.
  Assuming $\chi^{(2)}\sim 10^{-6}$ ESU ($\sim10^3$ pm/V), which is on the order of that for CdGeAs$_2$ crystals, and a control field of $I\sim~100$~kW focused on a spot of $D\sim 50$ $\mu$m  in diameter, one can estimate that the typical required value of the parameter $gL\sim1$ can be achieved for a slab thickness in the \emph{{microscopic}} range of $L\sim1$$ \mu$m, which is comparable with that of the multilayer~NIM samples fabricated to date.

\section{Quantum engineering of nonlinearity and coherent quantum control}\label{fwm}
\begin{figure}[h!]
\begin{center}
\resizebox{.127\columnwidth}{!}{
\includegraphics{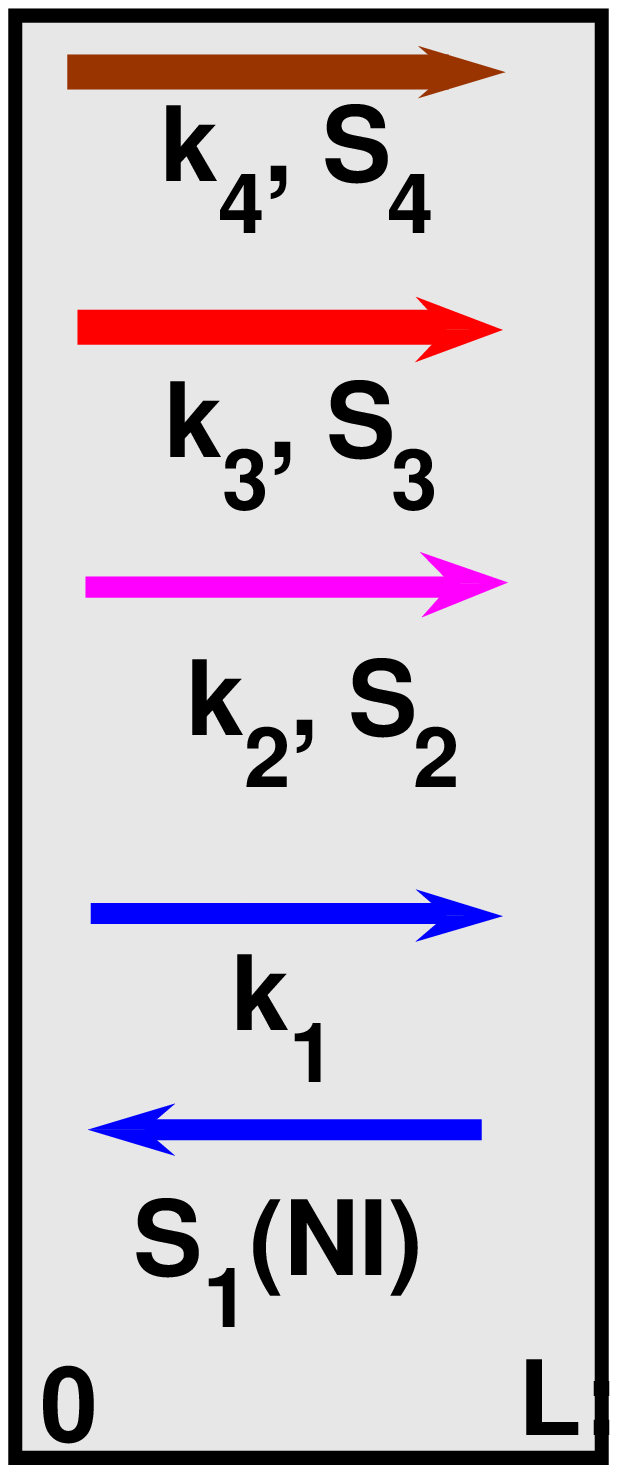}}
\resizebox{0.287\columnwidth}{!}{
\includegraphics{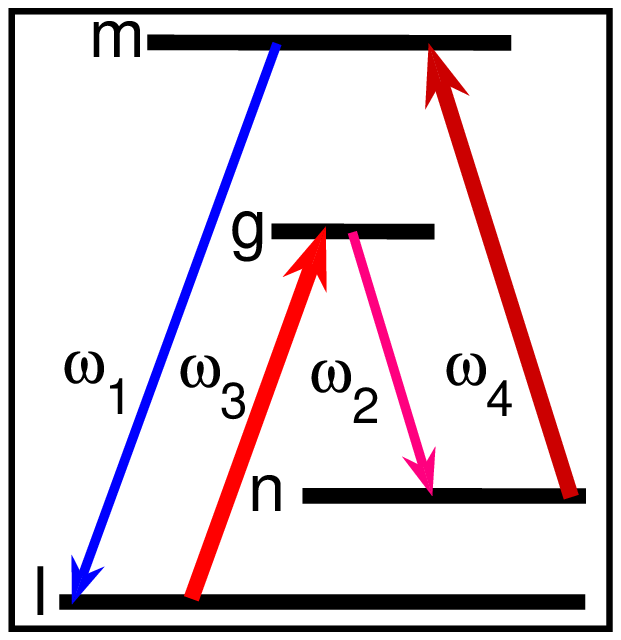}}
\resizebox{0.137\columnwidth}{!}{
\includegraphics{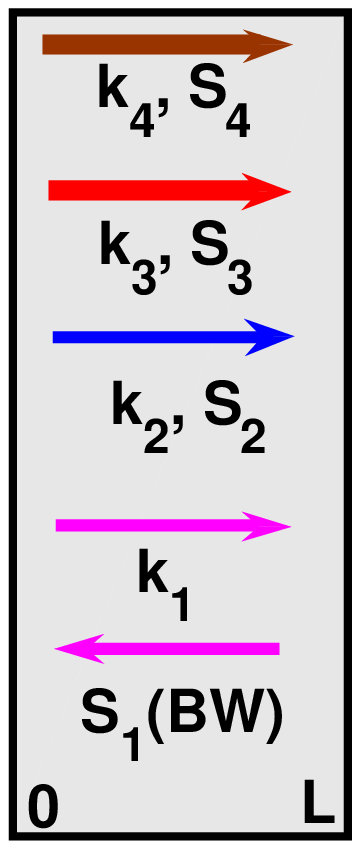}}
\resizebox{0.287\columnwidth}{!}{
\includegraphics{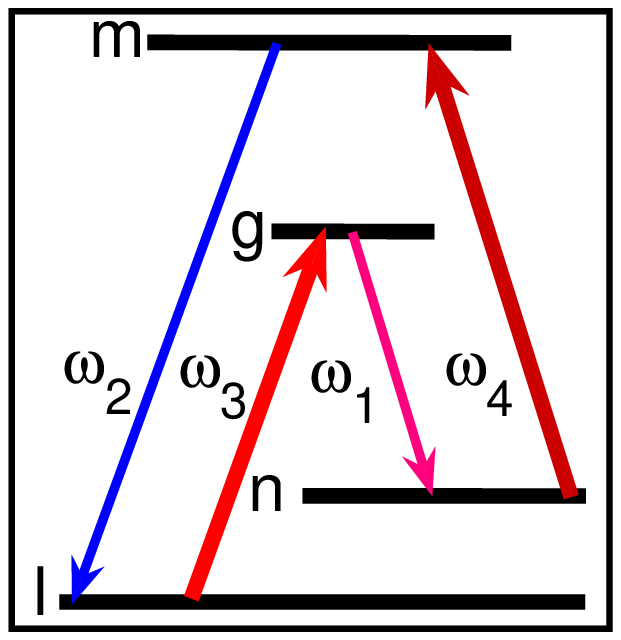}}\\
\hspace{-13mm}(a)\hspace{35mm} (b)\hspace{30mm} (c)\hspace{35mm} (d)
\caption{\small Coupling geometry and alternative schemes of four-wave mixing through embedded, resonant, nonlinear-optical centers. (a, c) Coupling geometry for four-wave mixing of backward and ordinary electromagnetic waves. $\mathbf{S}_1$, $\mathbf{k}_1$ and $\omega_1$ are the energy flux, wavevector and frequency for the backward-wave signal; $\mathbf{S}_2$, $\mathbf{k}_2$ and $\omega_2$ -- for the ordinary idler; $\mathbf{S}_{3,4}$, $\mathbf{k}_{3,4}$ and $\omega_{3,4}$ -- for the ordinary control fields. (b, d) Corresponding alternative schemes of quantum controlled four-wave mixing in embedded, resonant, nonlinear-optical centers with different ratios of the signal and idler absorption rates and nonlinear susceptibilities. (b) Shortest-wavelength negative-phase signal where, depending on the partial relaxation rates, parametric amplification can be assisted by the idler's  population-inversion or Raman-type amplification. (d) Longer-wavelength negative-phase signal where, depending on the partial relaxation rates, parametric amplification can be assisted by the signal's incoherent amplification attributed to population-inversion or Raman-type gain.
 } \label{fig6}
 \end{center}
\end{figure}

Figure~\ref{fig6} depicts the basic principles of independent quantum engineering of nonlinearity and its coherent quantum control.
Here, a resonantly enhanced, higher-order, electrical $ \chi ^{(3)}$  NLO response and four-wave mixing are employed in a composite metamaterial with embedded NLO centers (quantum dots, ions, or molecules). Two different options correspondingly depicted in panels (a), (b) and (c), (d) offer different means of quantum control by the driving fields at frequencies $\omega_3$ and $\omega_4$. Only the wave at $\omega_1$ is presumed to be a negative-index one, and all others are PI waves. Four-wave mixing $\omega_1=\omega_3+\omega_4-\omega_2$ replaces the three-wave mixing process described above, so that in this case two strong fields are used for controlling the process. Resonant processes at the depicted quantum transitions driven by the strong control fields make all linear and nonlinear characteristics of the composite intensity-dependent.
Among the specific features attributed to  the resonant coupling is the fact that the nonlinear susceptibility becomes complex in the vicinity of the introduced resonances, which is followed by a phase shift between the fields and the polarizations.  Due to the different contributions of the populations of the coupled energy levels to $\chi^{(3)}_1$ and to $\chi^{(3)}_2$, the latter becomes different and intensity-dependent \cite{GPRA}. The ratio of real and imaginary parts of $\chi^{(3)}_j$ varies with the change of the resonance offsets. It is also different for schemes depicted in Figs.~\ref{fig6}(b) and (d). It appears that the detrimental role of even a large phase mismatch can be diminished or eliminated through the appropriate adjustment of the real and imaginary parts of $\chi^{(3)}_j$. Such a possibility is not available for off-resonant wave mixing.
The properties of the coherent energy transfer from the control fields to the negative-phase signal depend strongly on the ratio of the absorption rates and NLO susceptibilities for the signal and the idler. Figures~\ref{fig6}(b) and (d) present two alternative options for quantum control of NLO propagation processes.  Figure~\ref{fig6}(b) depicts the scheme that offers the possibility of incoherent amplification (population inversion or Raman-type) for the idler \cite{OLM,APB09,OL09}. This possibility depends on the set of quantum relaxation rates inherent to the specific doping agent.
Figure~\ref{fig6}(d) presents a scheme that offers an alternative possibility. Here, the idler frequency is close to a higher-frequency transition from the ground state, and the signal corresponds to a lower-frequency transition between the excited states. Hence, absorption for the idler is typically larger than for the signal. This situation appears advantageous for robust transparency tailoring. No incoherent amplification is possible here for the idler, whereas incoherent absorption for the signal can be controlled and turned to amplification. Such a possibility is also contingent on the appropriate partial population and coherence relaxation rates attributed to the embedded centers. In all of the schemes outlined above, the linear and nonlinear local parameters can be tailored through quantum control by varying the intensities and frequency-resonance offsets for combinations of the two control driving fields.

Depending on the partial transition relaxation rates associated with levels \emph{m} and \emph{g}, the numerical model relevant to the coupling scheme depicted in Fig.~\ref{fig6}(d)  offers neither the possibility of population inversion nor Raman-type amplification. The fact that all involved optical transitions are absorptive determines  essentially different features of the overall resonant loss-compensation technique compared to that proposed in Refs. \cite{OLM,APB09,OL09} The following model, which is characteristic of ions and some molecules embedded in a solid host, has been adopted: energy level relaxation rates $\Gamma_n=20$, $\Gamma_g=\Gamma_m=120$ (all in $ 10^6$~s$^{-1}$); partial transition probabilities  $\gamma_{gn}=50$, $\gamma_{mn}=90$, (all in $ 10^6$~s$^{-1}$); homogeneous transition half-widths $\Gamma_{gl}$=1.8, $\Gamma_{mn}$=1.9, $\Gamma_{gn}$=1, $\Gamma_{ml}$=1.5, $\Gamma_{mg}=5\times10^{-2}$,
$\Gamma_{nl}=5\times10^{-3}$ (all in $10^{11}$ s$^{-1}$);
$\lambda_1=756$ nm and $\lambda_2=480$ nm.
The density-matrix method \cite{GPRA} has been used for calculating intensity-dependent local parameters while accounting for the quantum nonlinear interference effects. This allows one to investigate the changes in absorption, amplification, and refractive indices as well as in the magnitudes and signs of the NLO susceptibilities caused by the control fields. These changes depend on the population redistribution over the coupled levels, which in turn strongly depends on the ratio of the partial transition probabilities. Numerical simulations have led to promising conclusions regarding the enhanced reflectivity and amplification of the incident beam as well as possibilities of modulating these characteristics of the proposed microscopic metadevices \cite{OLM,OL09,APB09,JOA,WAS,EPJD,Ch}.

\subsection{Estimates}\label{est2}
For transition properties that are characteristic of molecules embedded in a solid environment, estimates show that the required intensity of the control fields at $\omega_3$ and $\omega_4$ are on the order of $I\sim$~1W/(0.1mm)$^2$. With a resonant absorption cross-section $\sigma_{40}~\sim~10^{-16}$~cm$^2$, which is typical for transitions with an oscillator strength of about one, and a concentration of embedded centers $ N~\sim~10^{19}$ cm$^{-3}$, one estimates $\alpha_{10}~\sim 10^3$~cm$^{-1}$ and the required slab thickness in the {microscopic} range $L~\sim (1 - 100) \mu$m. The contribution to the refractive index by the impurities then is estimated as $ \Delta n< 0.5(\lambda/4\pi)\alpha_{40}\sim 10^{-3}$, which essentially does not change the negative refractive index.
\section{Backward-wave optical phonons and distributed nonlinear-optical feedback}\label{ph}
A different scheme of TWM for ordinary and backward waves has been proposed previously \cite{ph}. It builds on stimulated Raman scattering (SRS) where two ordinary electromagnetic waves excite a backward, elastic, vibrational wave in a crystal and thus initiate TWM. The possibility of elastic BWs was predicted by L. I. Mandelstam in 1945 \cite{Ma}, who also pointed out that negative refraction is a general property of BWs. The basic idea of replacing the negative-index composites with ordinary crystals and mimicing the uncommon properties of coherent NLO energy exchange between the ordinary and backward waves was shown in Ref. \cite{ph} and is reviewed below.

\begin{figure}[h]
\begin{center}
\includegraphics[width=.7\columnwidth]{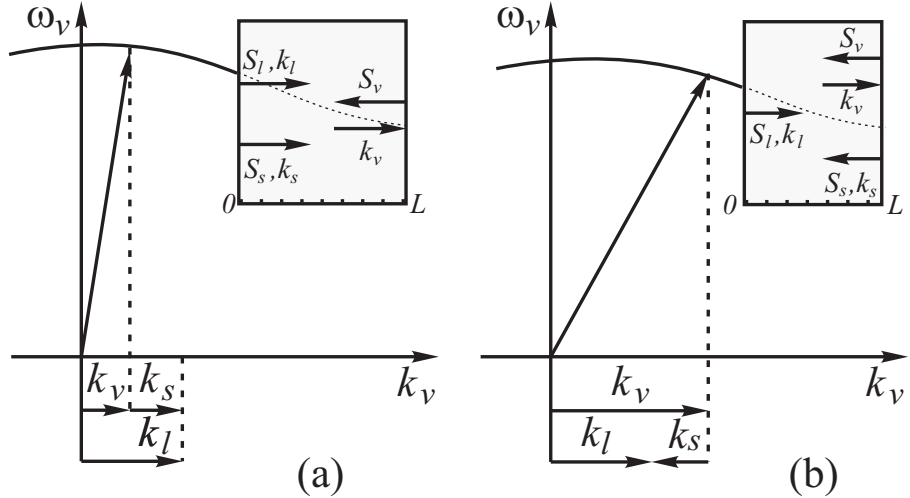}
\caption{\label{fig7} Negative dispersion of optical phonons and two phase-matching options: (a) co-propagating, and (b) contra-propagating fundamental, control, and Stokes signal waves. Insets: relative directions of the energy flows and the wavevectors.}
\end{center}
\end{figure}
The dispersion curve $\omega(k)$ of phonons in crystals containing more than one atom per unit cell has two branches: acoustic and optical. For the optical branch, the dispersion is negative in the range from zero to the boundary of the first Brillouin zone (Fig. \ref{fig7}). Hence, the group velocity of optical phonons, $\mathbf{v}_{v}^{gr}$, is antiparallel with respect to its wavevector, $\mathbf{k}_{v}^{ph}$, and phase velocity, $\mathbf{v}_{v}^{ph}$. This is because $v_{gr}=\partial \omega(k)/\partial k < 0$. Optical vibrations can be excited by light waves through two-photon (Raman) scattering. This gives us the grounds to consider these crystals as analogues to a medium with a negative refractive index at the phonon frequency and to examine the processes of parametric interaction of three waves. Two of these waves are ordinary electromagnetic waves. The third wave is the backward wave of elastic vibrations with directions of its energy flow and wavevector that are opposite to each other.
We investigat the extraordinary nonlinear propagation and output properties of the Stokes electromagnetic wave in one of two different coupling geometries, depicted in Fig. \ref{fig7}, with both utilizing the backward elastic waves. Such unusual properties are in contrast with those attributed to the counterparts in the standard schemes that build on the coupling of co-propagating photons and phonons. They are also different from the properties of the phase-matched mixing of optical and acoustic waves for the case when the latter has its energy flux and wavevector directed against those of one of the optical waves. The revealed properties can be utilized for the creation of optical switches, filters, amplifiers and cavity-free OPOs based on ordinary NLO crystals without the requirement of periodically poling at the nanoscale \cite{Kh}. Therefore, Fig.~\ref{fig7} depicts the concept of substituting a sophisticated, nanostructured, negative-index metal-dielectric composite by extensively used and studied ordinary crystals in order to mimic the  unparalleled properties of coherent NLO energy exchange between ordinary and backward waves.

\section{Conclusion}\label{con}

In conclusion, we have described the extraordinary properties of coherent, nonlinear-optical processes in negative-index metamaterials, such as second-harmonic generation, difference-frequency generation and optical parametric amplification. The feasibility of compensating the strong losses inherent to nanostructured, negative-index, plasmonic, metal-dielectric composites is outlined and illustrated with the aid of numerical simulations. The elimination of the detrimental role of optical losses in negative-index metamaterials is the key problem that limits the numerous revolutionary applications of this novel class of electromagnetic metamaterials. All-optical tailoring of
transparency and reflectivity is shown to be possible through coherent, nonlinear-optical energy transfer from the ordinary control wave to a negative-index, backward-wave field. This property is intrinsic to negative-index metamaterials. It is shown that besides the nonlinearity attributed to the building blocks of the negative-index host, a strong nonlinear optical response in the composite can be independently provided through embedded, resonant, four-level nonlinear-optical centers. This opens the way to independent nanoengineering and adjustment of the negative index and nonlinearities of metamaterials. In addition, we have described the opportunity for quantum control over the local optical parameters of the metamaterial in this case, which employs constructive and destructive quantum interference tailored by two auxiliary driving control fields. Such a possibility is proven with the aid of a realistic numerical model. The possibility of mimicing the extraordinary properties of coherent energy transfer between ordinary and the contra-propagating, backward waves inherent to negative-index, plasmonic metamaterials is described by making use of an easily available class of crystals with negative dispersion for optical phonons. Among the possible applications of the described nonlinear-optical processes are a novel class of miniature, all-optically tailored sensors, mirrors, frequency-tunable narrow-band filters, quantum switches, amplifiers, cavity-free microscopic optical parametric oscillators that allow generation of entangled counter-propagating left- and right-handed photons, and all-optical data-processing chips. The unique, unparalleled features of the underlying processes are outlined and include the strongly resonant behavior with respect to the material thickness, the density of the embedded resonant centers and the intensities of the control fields, the feasibility of negating the linear phase-mismatch introduced by the host material, and the role of absorption or, conversely, the supplementary nonparametric amplification of the idler.

\acknowledgments     
This work was supported in part by the U.S. National Science Foundation under Grant No. ECCS-1028353, by the U.S. Air Force Office of Scientific Research, by the U.S. Army Research Office under grants W911NF-0710261 and ARO-MURI grant W911NF-0910539, and by the U.S.  Office of Naval Research under  ONR MURI grant N00014-010942.
The authors thank S. A.  Myslivets for help with numerical simulations, thoughtful comments and useful discussions and Mark Thoreson  for help with preparing the manuscript.


\begin{thebibliography}{10}

\bibitem{Vesel1}   V. G. Veselago, ``Properties of materials having
simultaneously negative values of dielectric  and
magnetic susceptibilities,'' \emph{Sov. Phys. Solid State} \textbf{8}, 2854-2856 (1967).

\bibitem{Vesel2}  V.G. Veselago, ``The electrodynamics of substances with simultaneously negative values of epsilon and mu,'' [Sov. Phys. Usp. \textbf{10}, 509-514 (1968)] Usp. Fiz. Nauk \textbf{92}, 517-526 (1967).

\bibitem{Land}   L.D. Landau and E.L. Lifshits, [\emph{Electrodinamics of Continuous Media}], Ch. 9, 2nd. ed., Pergamon Press, New York (1960).

\bibitem{Jas} J. Valentine, S. Zhang, T. Zentgraf, E. Ulin-Avila, D. A. Genov, G. Bartal and X. Zhang,
``Three-dimensional optical metamaterial with a
negative refractive index,''
\emph{Nature} \textbf{455}, 376-378 (2008).

\bibitem{SM} C. M. Soukoulis and M. Wegener,
``Optical metamaterials — more bulky and less lossy,''
\emph{Science} \textbf{330}, 1633-1634 (2011).

\bibitem{Zh} N. I. Zheludev,
``A roadmap for metamaterials,''
\emph{OPN} \textbf{22}, 30-35 (2011).

\bibitem{Sh}
V.~M. {Shalaev}, ``{Optical negative-index metamaterials},'' { Nat.  Photonics}~{\bf 1}, 41-48 (2007).

\bibitem{Smith1} D.R. Smith, W. Padilla, D.C. Vier, S.C. Nemat-Nasser, and S. Shults,
``Composite Medium with Simultaneously Negative Permeability and
Permittivity," \emph{Phys. Rev. Lett}. \textbf{92}, 4184-4187 (2000).

\bibitem{Smith11} R. A. Shelby, D. R. Smith, and S. Schultz, ``Experimental
verification of a Negative Index of Refraction," \emph{Science}
\textbf{292}, 77-79  (2001).

\bibitem{Smith2} D.R. Smith, J. B. Pendry, and M. C. K. Wiltshire, ``Metamaterials and
Negative Refractive Index," \emph{Science} \textbf{305}, 788-790 (2004).

\bibitem{mu1} S. Linden, C. Enkrich, M. Wegener, J. Zhou, T. Koschny, and C. M. Soukoulis,  ``Magnetic Response of Metamaterials at 100
Terahertz," \emph{Science} \textbf{306},  1351-1353 (2004).

\bibitem{mu2} Z. Zhang, W. Fan, B. K. Minhas, A. Frauenglass, K. J. Malloy,
and S. R. J. Brueck, ``Midinfrared Resonant Magnetic Nanostructures
Exhibiting a Negative Permeability,"  \emph{Phys. Rev. Lett}. \textbf{94},
 037402 (2005).

\bibitem{mu3} A. N. Grigorenko, A. K. Geim, N. F. Gleeson, Y. Zhang, A. A.
Firsov, I. Y. Khrushchev, and J. Petrovic, ``Nanofabricated media
with negative permeability at visible frequencies,"  \emph{Nature},
\textbf{438}, 335-337 (2005).

\bibitem{NIMExp1} V. M. Shalaev, W. Cai, U. Chettiar, H.-K. Yuan, A. K. Sarychev, V. P. Drachev, and A. V. Kildishev, ``Negative index of refraction in optical metamaterials," \emph{Optics Letters} \textbf{30}, 3356-3358 (2005).

\bibitem{NIMExp2} S. Zhang, W. Fan, N. C. Panoiu, K. J. Malloy, R. M. Osgood, and S. R. J. Brueck, ``Experimental Demonstration of
Near-Infrared Negative-Index Metamaterials," \emph{Phys. Rev. Lett}.
\textbf{95}, 137404 (2005).

\bibitem{NIMExp21} S. Zhang, W. Fan, K. J. Malloy, S. R. J. Brueck,
N. C. Panoiu, and R. M. Osgood, ``Demonstration of metal-dielectric
negative-index metamaterials with improved performance at optical
frequencies," \emph{J. Opt. Soc. Am.} B \textbf{23} , 434-438 (2006).


\bibitem{WeLi} M. Wegener and S. Linden, ``Giving light yet another new twist,'' \emph{Physics} \textbf{2}, 3 (2009).

\bibitem{Pl} E. Plum, J. Zhou, J. Dong, V. A. Fedotov, T. Koschny, C. M. Soukoulis, and N. I. Zheludev, ``Metamaterial with negative index due to chirality,'' \emph{Phys. Rev}. \textbf{B 79}, 035407 (2009).

\bibitem{SZh} S. Zhang, Y.-S. Park, J. Li, X. Lu, W. Zhang, and X. Zhang, ``Negative Refractive Index in Chiral Metamaterials,''
\emph{Phys. Rev. Lett}. \textbf{102}, 023901 (2009).

\bibitem{caus1} R. W. Ziolkowski and E. Heyman, "Wave Propagation in Media Having
Negative Permittivity and Permeability," \emph{Phys. Rev}. E \textbf{64},
056625 (15) (2001).

\bibitem{caus2} R.W. Ziolkowski, and A. Kipple, ``Causality and Double-Negative
Metamaterials," \emph{Phys. Rev}. E \textbf{68}, 026615 (2003).

\bibitem{amp1}  S. A. Ramakrishna and J. B. Pendry, ``Removal of Absorption
and Increase in Resolution in a Near-Field Lens via Optical Gain,"
\emph{Phys. Rev.} B \textbf{67}, 201101 (2003).

\bibitem{amp2} T. A. Klar, A. V. Kildyshev, V. P. Drachev, and V.M. Shalaev,
``Negative-Index Metamterials: Going Optical,"  \emph{IEEE Journal of Selected Topics in Quantum Electronics} \textbf{12}, 1106-1115 (2006).

\bibitem{amp3} U. K. Chettiar, A. V. Kildishev, T. A. Klar, H. -K. Yuan, W. Cai, A. K. Sarychev, V. P. Drachev, and V. M. Shalaev, ``From Low-loss to Lossless Optical Negative-Index Materials," \emph{CLEO/QELS-06 Annual Meeting Proceedings}, Long Beach, CA, May 21-26, (2006).

\bibitem{amp4} A. Sarychev and G. Tartakovsky, ''Magnetic plasmonic metamaterials in actively pumped host medium and plasmonic nanolaser,'' \emph{Phys. Rev}. B \textbf{75}, 085436 (2007).

\bibitem{amp5} A. N. Lagarkov, A. K. Sarychev, V. N. Kissel, G. Tartakovsky, ``Superresolution and enhancement in metamaterials'', \emph{Physics-Uspekhi} \textbf{52} (9), 959–967 (2009).

\bibitem{ampr} R. Espinola, J. Dadap, R. Osgood, Jr., Sh. McNab, and Yu. Vlasov,
``Raman Amplification in Ultrasmall Silicon-on-Insulator Wire
Waveguides," \emph{Opt. Express} \textbf{12}, 3713-18 (2004).

\bibitem{Siv} Yonatan Sivan, Shumin Xiao, Uday K. Chettiar, Alexander V. Kildishev, and Vladimir M. Shalaev, ``Frequency-domain simulations of a negative-index material with embedded gain,'' \emph{Optics Express} \textbf{17},  24060-24074 (2009).

\bibitem{Nog} M. A. Noginov, G. Zhu, A. M. Belgrave, R. Bakker, V. M. Shalaev, E. E. Narimanov, S. Stout, E. Herz, T. Suteewong and U. Wiesner, ``Demonstration of a spaser-based nanolaser,'' \emph{Nature}  \textbf{460}, 1110-1114 (2009).

\bibitem{Ampl}  Xiao, S.,  Drachev, V.P.,  Kildishev, A.V.,  Ni, X., Chettiar, U.K.,  Yuan, H.-K. and  Shalaev, V.M., 2010, ``Loss-free and active optical negative-index metamaterials," \emph{Nature},  \textbf{ 466 (7307)},   735-738 (2010).

\bibitem{Lap1}  M. Lapine, M. Gorkunov, and K. H. Ringhofer,
``Nonlinearity of a metamaterial arising from diode insertions into
resonant conductive elements," \emph{Phys. Rev}. E \textbf{67}, 065601
(2003).

\bibitem{Zhar}  A. A. Zharov, I. V. Shadrivov, and Yu. S. Kivshar,
``Nonlinear properties of left-handed metamaterials," \emph{Phys. Rev. Lett}. \textbf{91}, 037401 (2003).

\bibitem{Lap2}  M. Lapine and M. Gorkunov, ``Three-wave Coupling of
Microwaves in Metamaterials with Nonlinear Resonant Condactive
Elements," \emph{Phys. Rev}. E \textbf{70}, 66601 (2004).

\bibitem{Gorkopa} M. V. Gorkunov,  Il.V. Shadrivov, and Yu. S.
Kivshar, ``Enhanced Parametric Processes In Binary Metamaterials,"
\emph{Appl. Phys. Lett}. \textbf{87},  071912-3 (2005).

\bibitem{Shadr} I. V. Shadrivov, S. K. Morrison, and Yu. S. Kivshar, ``Tunable
Split-Ring Resonators for Nonlinear Negative-Index Metamaterials,"
\emph{Optics Express}  \textbf{14}, 9344-9349 (2006).

\bibitem{shg1e} M. W. Klein, C. Enkrich, M. Wegener, and S. Linden,
``Second-Harmonic Generation from Magnetic Metamaterials," \emph{Science}
\textbf{313}, 502-504 (2006).

\bibitem{shg2e} M. W. Klein, C. Enkrich, M. Wegener, J. F\"orstner,
J. V. Moloney, W. Hoyer, T Stroucken, T Meyer, S. W. Koch, and S.
Linden, ``Optical Experiments on Second-Harmonic Generation with
Metamaterials Composed of Split-Ring Resonators," Paper OThE3, in  [\emph{CLEO/QELS
Proceedings}],  Long Beach, California,   May 21-26 (2006).
also Paper TuC5 in [\emph{Proceedins of Photonic Metamaterials: From Random to Periodic (META)}], Grand Bahama Island, The Bahamas, June 5-8 (2006).

\bibitem{Kl}
M. W. Klein, M. Wegener, N. Feth,
and S. Linden,
``Experiments on
second- and third-harmonic generation from magnetic
metamaterials,''
\emph{Opt. Express} \textbf{15},  5238-5247 (2007);
erratum:\emph{ibid}, \textbf{16}, 8055 (2008).

\bibitem{Shad}	I. V. Shadrivov, A. B. Kozyrev, D. W. van der Weide and Y.S. Kivshar,``Tunable transmission and harmonic generation in nonlinear metamaterials," \emph{Appl. Phys. Let}. \textbf{93}, 161903 (2008).
\bibitem{Pet}	J. Petschulat, A. Chipouline, A. T\"{u}nnermann, T. Pertsch, C. Menzel, C. Rockstuhl and F. Lederer, ``Multipole nonlinearity of metamaterials," \emph{Phys. Rev}. A \textbf{80}, 063828-7 (2009);
\bibitem{Ze}	Y. Zeng and J. V. Moloney, ``Volume electric dipole origin of second-harmonic generation from metallic membrane with noncentrosymmetric patterns," \emph{Opt. Lett}. \textbf{34}, 2844-46 (2009).
\bibitem{Nie}	F. B. P. Niesler, N. Feth, S. Linden, J. Niegemann, J. Gieseler, K. Busch and M. Wegener, ``Second-harmonic generation from split-ring resonators on a GaAs substrate," \emph{Opt. Lett}. \textbf{34}, 1997-99 (2009).

\bibitem{Agr} V. M. Agranovich, Y. R. Shen, R. H. Baughman, and A. A. Zakhidov,
``Linear and Nonlinear Wave Propagation in Negative Refraction
Metamaterials," \emph{Phys. Rev}. B \textbf{69}, 165112 (2004).

\bibitem{Lens} A. A. Zharov, N. A. Zharova, I. V. Shadrivov, and Yu.S. Kivshar, ``Subwavelength Imaging with Opaque Nonlinear Left-Handed
Lenses," \emph{Appl. Phys. Lett}. \textbf{88}, 091104-3 (2005).

\bibitem{Lens1} N. A. Zharova, I. V. Shadrivov, A. A. Zharov, and Yu. S. Kivshar,
``Nonlinear Transmission and Spatiotemporal Solitons in Metamaterials
with Negative Refraction," \emph{Optics Express} \textbf{13}, 1291 (2005).

\bibitem{KivSHG} I. V. Shadrivov, A. A. Zharov  and Yu. S. Kivshar, ``Second-harmonic generation in nonlinear left-handed metamaterials,'' \emph{J. Opt. Soc. Am. B}, \textbf{23}, 529-534 (2006).

\bibitem{Kozyr} A. B. Kozyrev, H. Kim, A. Karbassi, and D. W. van der Weide, ``Waver Propagation in Nonlinear Left-handed Transmission line
Media,"  \emph{Appl. Phys. Lett}. \textbf{87}, 121109-3 (2005).


\bibitem{Sc} M. Scalora, G.  D'Aguanno, M. Bloemer,  M. Centini, N. Mattiucci, D. de Ceglia  and   Yu. S. Kivshar,   ``Dynamics of Short Pulses and  Phase Matched Second Harmonic Generation in Negative
Index Materials," \emph{Opt. Express}, \textbf{ 14}, 4746-56 (2006).

\bibitem{SHG}
A. K. {Popov}, V. V.  {Slabko}  and V. M. {Shalaev},  ``{Second harmonic  generation in left-handed metamaterials},''  {\em Laser Phys. Lett.}~\textbf{3}, 293-296 (2006).

\bibitem{APB}
A. K. Popov and V. M. Shalaev,  ``{Negative-index metamaterials:
  second-harmonic generation, Manley Rowe relations and parametric
  amplification},'' {\em Applied Physics B: Lasers and Optics}~{ \textbf{84}},   131-137 (2006).

\bibitem{met} A. K. Popov and V. M. Shalaev, ``Second Harmonic Generation and Parametric
Amplification in Negative Index Metamaterials," in |\emph{Proceedins of
Photonic Metamaterials: From Random to Periodic (META)}|, Grand Bahama Island, The Bahamas, June 5-8 (2006), Paper TuC4.

\bibitem{OL} A. K. Popov and V. M. Shalaev, ``Compensating losses in negative-index metamaterials by optical parametric amplification,''  \emph{Opt. Lett}., \textbf{31}, 2169-2172 (2006).

\bibitem{OLM}
A.~K. {Popov}, S.~A. {Myslivets}, T.~F. {George}, and V.~M. {Shalaev},   ``{Four-wave mixing, quantum control, and compensating losses in doped  negative-index photonic metamaterials},''\emph{ Opt. Lett.} ~{ \textbf{32}},   3044-3046 (2007).

\bibitem{APL}
 A.~K. Popov and S.~A. Myslivets, ``Transformable broad-band transparency and  amplification in negative-index films,'' {\em Applied Physics Letters}~{ \textbf{93}}, 191117(3) (2008).

 \bibitem{LSh}  N. M. Litchinitser and V. M. Shalaev, ``Loss as a route to transparency,"  \emph{Nat. Photonics},  \textbf{ 3}, 75-79 (2009).

\bibitem{APB09}
A.~K. {Popov}, S.~A. {Myslivets}, and V.~M. {Shalaev}, ``{Resonant nonlinear  optics of backward waves in negative-index metamaterials},'' \emph{ Applied Physics B: Lasers and Optics}~ \textbf{96}, 315-323 (2009).

\bibitem{OL09}
 A.~K. {Popov}, S.~A. {Myslivets}, and V.~M. {Shalaev},  ``Microscopic mirrorless negative-index optical parametric oscillator, ''\emph{ Opt. Lett}. \textbf{ 34},  1165-1167 (2009).

\bibitem{SPI}	A. K. Popov, S. A. Myslivets and V.M. Shalaev, ``Plasmonics: nonlinear optics, negative phase and transformable transparency'' (Invited Paper), in |\emph{Plasmonics: Nanoimaging, Nanofabrication, and their Applications V}|,  Satoshi Kawata, Vladimir M. Shalaev, Din Ping Tsai, eds., \emph{Proc. SPIE} \textbf{7395}, 73950Z-1(12) (2009).

\bibitem{JOA} A. K. Popov, S. A. Myslivets, and V. M. Shalaev, ``Coherent nonlinear optics and quantum control in negative-index metamaterials,'' \emph{J. Opt. A: Pure Appl. Opt.} \textbf{11}, 114028(13) (2009).

\bibitem{WAS}  A.~K. Popov and S.~A. Myslivets, ``Numerical Simulations of  Negative-Index Nanocomposites and Backward-Wave Photonic Microdevices,'' {\em  "ICMS 2010 : International Conference on Modeling and Simulation," Proc.  WASET}~\textbf{  61}, 107-121 (2010).

    \bibitem{EPJD}	 A. K. Popov,   ``Nonlinear optics of backward waves and extraordinary features of plasmonic nonlinear-optical microdevices,''  {\emph{Eur. Phys. J. D} } \textbf{ 58},  263-274  (2010) (Topical issue on {Laser Dynamics and Nonlinear Photonics}).
\bibitem{Ch}	 A. K. Popov  and  T. F. George,  ``Computational Studies of Tailored Negative-Index Metamaterials and Microdevices, ''  Chapter 13, in [\emph{Computational Studies of New Materials II: From Ultrafast Processes and Nanostructures to Optoelectronics, Energy Storage and Nanomedicine}],  T. F. George, D. Jelski, R. R. Letfullin and G. Zhang, eds., World Scientific, Singapore, (2011), pp. 331-377.

 \bibitem{Laz}  N. Lazarides and GP Tsironis,
``Coupled nonlinear Schr\"{o}dinger field equations for electromagnetic wave propagation in nonlinear left-handed materials," \emph{Phys. Rev}. E \textbf{71}, 036614-1-4 (2005).

\bibitem{Tas} P. Tassin, L. Gelens, J. Danckaert, I. Veretennicoff, G. Van der Sande, P. Kockaert, and M. Tlidi,
    ``Dissipative structures in left-handed material cavity optics,"
    \emph{Chaos} \textbf{17}, 037116-1-11 (2007).

\bibitem{Kos} P. Kockaert, P. Tassin, G. Van der Sande, I. Veretennicoff, and M. Tlidi,
``Negative diffraction pattern dynamics in nonlinear cavities with left-handed materials,"
\emph{Phys. Rev}. A \textbf{74}, 033822-1-8 (2006).

\bibitem{Boa} A. D. Boardman, N. King, R. C. Mitchell-Thomas, V. N. Malnev, Y. G. Rapoport,
``Gain control and diffraction-managed solitons in metamaterials,"
\emph{Metamaterials} \textbf{2}, 145-154 (2008).

\bibitem{Agu}   G. D'Aguanno, N. Mattiucci, M. Scalora, and M. J Bloemer,
   ``Bright and Dark Gap Solitons in a Negative Index Fabry-Perot Etalon,"
    \emph{Phys. Rev. Lett}. \textbf{93}, 213902(1) (2004).

\bibitem{Lit}
N. M. Litchinitser,  I. R. Gabitov, A. I. Maimistov,
and V. M. Shalaev,
``Negative Refractive Index Metamaterials in Optics,"
 \emph{{ Progress in Optics}},  \textbf{51},
 1-68 (2007).

\bibitem{LitOL}
N. M. Litchinitser,  I. R. Gabitov, A. I. Maimistov,
and V. M. Shalaev,
``Effect of an optical negative index thin film on optical
bistability,"
\emph{Opt. Lett.} \textbf{32}, 151-153 (2007).
%

\bibitem{Gab} A. I. Maimistov and I. R. Gabitov,
``Nonlinear optical effects in artificial materials,"
\emph{ Eur. Phys. J.}
Special Topics,
\textbf{147}, 265-286 (2007).

\bibitem{El} S. O. Elyutin, A. I.  Maimistov,  and I. R.  Gabitov, ``On the third harmonic generation in a medium with negative pump wave refraction,'' \emph{JETP}, \textbf{111}, 157-169 (2010).
\bibitem{Pas}
C. Canalias and V. ~Pasiskevicius,
``Mirrorless
optical parametric oscillator,"
\emph{ Nat. Photonics} \textbf{1},  459-462 (2007).

\bibitem{Kh} J.~B. Khurgin,
``Mirrorless magic,"
\emph{Nat. Photonics} \textbf{1}, 446-447 (2007) (and references therein).

\bibitem{Har}
S. E. Harris,
``Proposed backward wave oscillations in the
infrared,"
\emph{ Appl. Phys. Lett.} \textbf{9}, 114-117 (1966).

\bibitem{Vol} K. I. Volyak, A. S. Gorshkov,
``Investigations of a
reverse-wave parametric oscillator,"
\emph{Radiotekhnika i Elektronika (Radiotechnics and Electronics)} \textbf{18}
2075-2082 (1973).

\bibitem{Yar} A. Yariv, { [Quantum Electronics]}, 2nd Ed.,
 Wiley, New York  (1975), Ch. 18.

\bibitem{MR56}
J.~M. Manley and H.~E. Rowe, ``Some general properties of nonlinear
  elements--{P}art {I}. {G}eneral energy
  relations," \emph{Proc. IRE}, \textbf{44},   904-913
   (1956).

\bibitem{MR} J. M. Manley and H. E. Rowe,
``General energy relations in nonlinear reactances,"
\emph{Proc. IRE}
\textbf{47}, 2115-2116 (1959).

\bibitem{GPRA} A. K. Popov, S. A. Myslivets, and T. F. George,
``Nonlinear interference effects and all-optical switching in
optically-dense inhomogeneously-broadened media,''
\emph{Phys. Rev}. A \textbf{71}, 043811(13) (2005).


\bibitem{Ma}  L. I. Mandelstam,
Group velocity in a crystall lattice,
 \emph{ZhETF}  \textbf{{15}}, 475-478 (1945).

\bibitem{ph}  V. V. Slabko,  S. A.  Myslivets,  M. I. Shalaev  and A. K. Popov,   ``Negative group velocity and three-wave mixing in dielectric crystals,'' \emph{arXiv}: 1104.0891 v1. (2011).

\end{thebibliography}
\end{document}